\documentclass[10pt,twocolumn,letterpaper]{article}

\usepackage[pagenumbers]{cvpr} 

\usepackage{booktabs} 
\usepackage{indentfirst} 
\usepackage{threeparttable} 
\usepackage{multirow} 
\usepackage{makecell} 
\usepackage{amsmath}
\usepackage{bm}
\definecolor{cvprblue}{rgb}{0.21,0.49,0.74}
\usepackage[pagebackref,breaklinks,colorlinks=true, linkcolor=red, citecolor=cvprblue]{hyperref}

\def\paperID{10325} 
\def\confName{CVPR}
\def\confYear{2025}

\title{Augmented Deep Contexts for Spatially Embedded Video Coding}

\author{Yifan Bian\quad Chuanbo Tang\quad Li Li\quad Dong Liu
\thanks{This work was supported by the Natural Science Foundation of China under Grants 62036005 and 62021001. 
We acknowledge the support of GPU cluster built by MCC Lab of Information Science and Technology Institution, USTC.
\emph{(Corresponding author: Dong Liu.)}}\\
{\small MOE Key Laboratory of Brain-Inspired Intelligent Perception and Cognition}\\
{\small University of Science and Technology of China, Hefei 230027, China}\\
{\tt\small togelbian@gmail.com, cbtang@mail.ustc.edu.cn, \{lil1,dongeliu\}@ustc.edu.cn}
}

\makeatletter
\def\thanks#1{\protected@xdef\@thanks{\@thanks
        \protect\footnotetext{#1}}}
\makeatother

\begin{document}
\maketitle
\begin{abstract}
	\indent Most Neural Video Codecs (NVCs) only employ temporal references to generate temporal-only contexts and latent prior.
	These temporal-only NVCs fail to handle large motions or emerging objects due to limited contexts and misaligned latent prior.
	To relieve the limitations, we propose a Spatially Embedded Video Codec (SEVC), 
	in which the low-resolution video is compressed for spatial references.
	Firstly, our SEVC leverages both spatial and temporal references to generate augmented motion vectors and hybrid spatial-temporal contexts.
	Secondly, to address the misalignment issue in latent prior and enrich the prior information,
	we introduce a spatial-guided latent prior augmented by multiple temporal latent representations.
	At last, we design a joint spatial-temporal optimization to learn quality-adaptive bit allocation for spatial references, further boosting rate-distortion performance.
	Experimental results show that our SEVC effectively alleviates the limitations in handling large motions or emerging objects,
	and also reduces 11.9\% more bitrate than the previous state-of-the-art NVC while providing an additional low-resolution bitstream.
	Our code and model are available at \url{https://github.com/EsakaK/SEVC}.
\end{abstract}
    
\section{Introduction}
\label{sec:intro}

\begin{figure}[htb]
	\centering
	\includegraphics[width=0.9\linewidth]{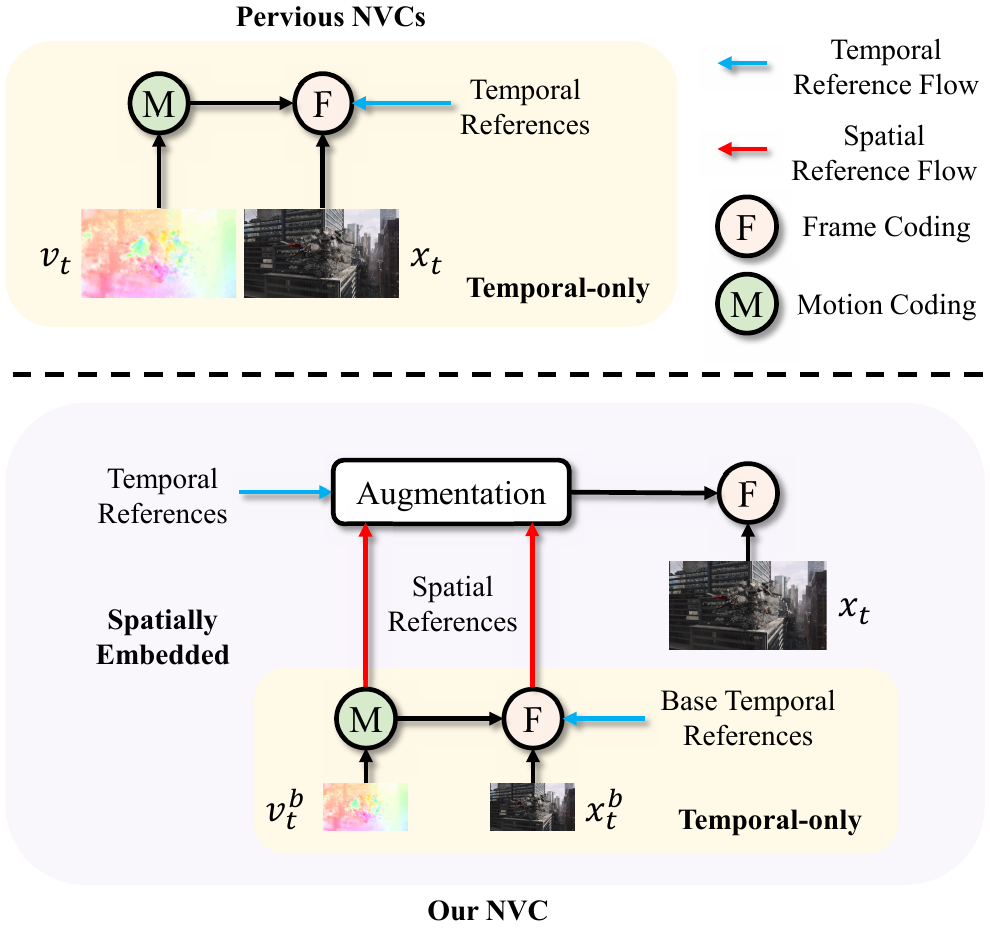}
	\vspace{-0.5em}
	\caption{
		Compared with previous Neural Video Codecs (NVCs), 
		our proposed codec embeds a base temporal-only codec to compress the low-resolution motion $v_t^b$ and frame $x_t^b$ and provide additional spatial references.
		Spatial and temporal references then augment deep contexts for frame coding.
	}
	\vspace{-1.5em}
	\label{fig:coding}
\end{figure}

The fundamental problem in video coding lies in determining rich references and effectively utilizing them to obtain an accurate prediction.
The richer the references, the better the prediction, and the more bitrate saved.\par

Over the years, advancements in traditional video codecs~\cite{h261,h.263,h.264,h.265,h.266} have heavily relied on more efficient utilization of temporal references.
For example, more complex motion models~\cite{wiegand2005affine, li2024uniformly, mesh1,li2024object},
are proposed to improve the quality of prediction.
Inspired by them, recent developments in neural video codecs (NVCs)~\cite{tcm, fvc, dc, tang2024offline, ssd,li2024neural} have also concentrated on
improving the utilization efficiency of temporal references.\par

However, for most recent NVCs, employing only temporal references fails to handle large motions or emerging objects.
One reason is that the motion models in existing NVCs are not strong enough to accurately estimate motion vectors (MVs)~\cite{uncertainty}, 
and inevitable motion coding errors exacerbate this inaccuracy.
Inaccurate MVs disrupt the context mining, resulting in poor temporal contexts.
Another reason is that these temporal-only contexts rely on limited temporal references, 
such as a single previously decoded feature and the misaligned latent prior~\cite{hem}, 
challenging to describe emerging objects.
Therefore, mitigating the MV errors and generating richer contexts still represent the limitations encountered by NVCs.

\begin{figure*}[tb]
	\centering
	\includegraphics[width=1.0\linewidth]{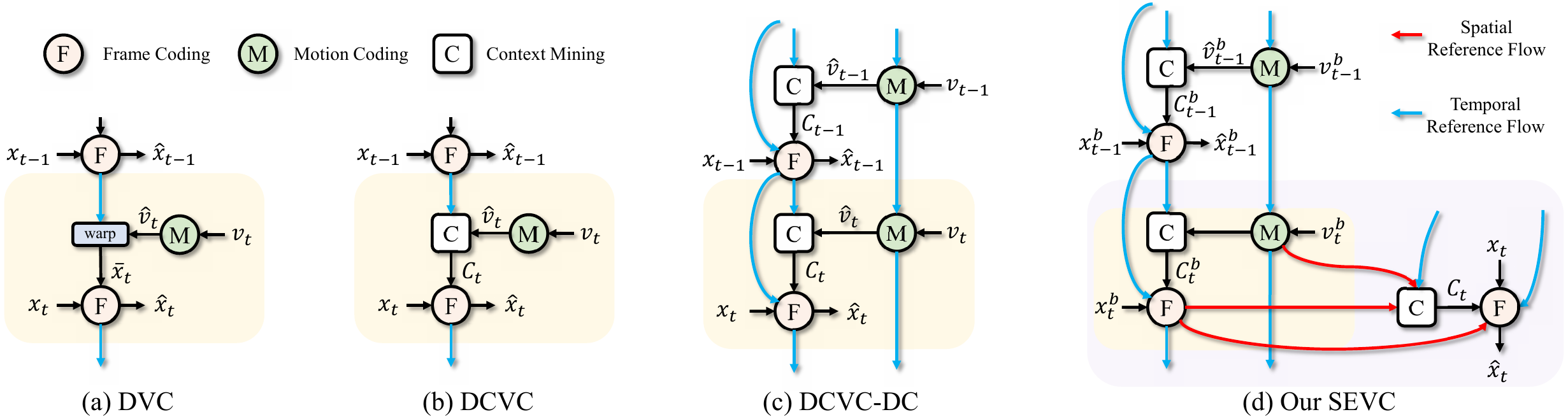}
	\vspace{-1.5em}
	\caption{
		Comparison of our Spatially Embedded Video Codec (SEVC) and previous NVCs.
		(a) compresses the residual between the prediction frame $\bar{x}_t$ and the input frame $x_t$.
		(b) extracts deep contexts $C_t$ from the previously reconstructed frame to serve as the coding condition.
		(c) further improves the utilization of temporal references in both motion coding and frame coding.
		(d) conducts a base coding on the low-resolution input frame $x_t^b$ and utilizes both spatial and temporal references for frame coding.
	}
	\vspace{-1.5em}
	\label{fig:fra_com}
\end{figure*}

Deviating from previous temporal-only NVCs,
we draw inspiration from Video Super-Resolution (VSR)~\cite{shi2022rethinking, Kappeler2016,chan2021basicvsr,chan2022basicvsr++}
by incorporating additional low-resolution spatial references to reconstruct the high-resolution frame.
To alleviate the limitations caused by insufficient references, we propose our \textbf{S}patially \textbf{E}mbedded \textbf{V}ideo \textbf{C}odec (SEVC).
As shown in Figure~\ref{fig:coding}, in previous NVCs, only temporal references are utilized for frame coding.
By contrast, our SEVC embeds a base temporal-only codec~\cite{dc} to provide additional spatial references.
Then spatial references and temporal references are augmented for frame coding.\par

A more detailed comparison is shown in Figure~\ref{fig:fra_com}.
DVC~\cite{dvc} considers the aligned reconstructed frame as the prediction and the residual between the prediction and input frame is compressed. 
DCVC~\cite{dcvc} proposes replacing the prediction frame with deep contexts, which serves as the condition in frame coding.
DCVC-DC~\cite{dc} further improves the utilization of temporal references in both motion coding and frame coding.
However, these NVCs still rely on temporal-limited references.\par

Regarding DCVC-DC~\cite{dc} as our base codec, we mainly focus on designing the augmentative codec.
Specifically, our SEVC first compresses a 4x downsampling low-resolution frame, resulting in three low-resolution spatial references~\cite{bian2024lssvc}: the base MVs, the spatial feature, and the spatial latent representation.
Then a Motion and Feature Co-Augmentation (MFCA) module is proposed to augment base MVs and the spatial feature.
The MFCA leverages the temporal feature to
progressively amplify the quality of base MVs and the spatial feature.
With the augmented base MVs and the spatial feature, hybrid spatial-temporal deep contexts are obtained for better frame coding.
In addition to generating augmented deep contexts,
introducing spatial references also benefits the entropy model.
Given spatial latent representation as the guidance,
multiple temporal latent representations could be aligned 
implicitly via Transformers~\cite{shi2022rethinking,liu2021swin}.
To further boost rate-distortion (RD) performance,
we introduce a joint spatial-temporal optimization strategy to learn better bit allocation for spatial references.\par

Experiments show that our proposed SEVC effectively alleviates the coding limitations encountered by previous NVCs in handling large motions or emerging objects.
When evaluated on commonly used test sets, our SEVC achieves state-of-the-art (SOTA) performance and far surpasses previous NVCs~\cite{ho2022canf, tcm, hem,dc, li2024neural, ssd} in sequences with aforementioned conditions.
Our contributions are summarized as follows:
\begin{itemize}
    \item We introduce embedding a base codec for additional spatial references to alleviate limitations in handling large motions and emerging objects. Spatial references benefit the generation of contexts and latent prior.
	\item For augmenting contexts, we propose a Motion and Feature Co-Augmentation module that progressively generates hybrid spatial-temporal contexts. 
	For augmenting the latent prior, we introduce a spatial-guided latent prior that merges multiple temporal latent representations.
	\item We develop a joint spatial-temporal optimization strategy that helps the spatially embedded codec learn better bit allocation, further boosting RD performance.
	\item Our proposed codec surpasses the previous SOTA NVC by 11.9\% more bitrate saving while providing a low-resolution bitstream.
\end{itemize}
\section{Related Work}
\label{sec:related}

\subsection{Neural Video Coding}
Recent advancements in Neural Video Codecs (NVCs) have demonstrated the
the superior potential of neural video coding.
Existing studies on NVCs can be broadly categorized into two types: residual coding and conditional coding.
Residual coding~\cite{res_1, res_2, res_3, res_4, dvc, dvc_pro, fvc, hu2022coarse, agustsson2020scale} addresses the frame differences to reduce redundancy,
while conditional coding~\cite{cond_1, cond_2, dcvc, tcm, ho2022canf, hem, dc, li2024neural, jiang2024sparse} exploits correlation based on
contextual information from temporal references.
Another kind of NVCs based on neural video representations~\cite{chen2021nerv,chen2023hnerv,zhao2023dnerv,li2022nerv} is also gaining attention, but we will not discuss them because the technical routes are so far apart.
Our SEVC adopts conditional coding due to its flexible use of conditions but explores spatial references to generate richer contextual information for frame coding.

\subsection{Deep Contexts Augmentation}
Conditional coding exploits contextual information as the prediction,
where the richness of contexts determines how good the prediction is.
Most existing NVCs augment contexts~\cite{dcvc, tcm, ssd, dc,mentzer2022vct,alexandre2023hierarchical} through better utilization of temporal references.
DCVC-TCM~\cite{tcm} incorporates a Temporal Context Mining module that extracts multi-scale contexts from propagated features instead of decoded frames.
DCVC-DC~\cite{dc} proposes the Group-Based Offset Diversity Alignment to estimate offset maps for different feature groups,
augmenting the capacity of temporal contexts to describe diverse scenarios.
In the entropy model, Mentzer \textit{et al.} use Transformers~\cite{vaswani2017attention}
to directly predict the contexts in latent space.
DCVC-HEM~\cite{hem} introduces the temporal latent prior\footnote{To distinguish contexts in feature space and latent space, the term ``contexts'' 
refers to contexts in feature space, while ``prior'' denotes contexts in latent space.} to further exploit temporal correlation.\par

However, temporal contexts are insufficient to describe newly appearing objects and are vulnerable to the accuracy of MVs~\cite{uncertainty},
thereby failing to describe large motions or emerging objects.
Meanwhile, there are works~\cite{spatial_ref_1,alexandre2023hierarchical} that leverages spatial references.
Yang \textit{et al.}~\cite{spatial_ref_1} compress the space-time down-sampled video and reconstruct it through a space-time super-resolution network.
Alexandre \textit{et al.}~\cite{alexandre2023hierarchical} interpolate the prediction of the input low-resolution frame with the properties of random-access scenarios,
and a super-resolution network is used to get a high-resolution prediction frame.
These methods are not oriented to low-delay scenarios, where the interpolated MVs are unavailable.
In addition, they consider prediction in pixel space which is inferior to contexts in feature space~\cite{dcvc,fvc}.
By contrast, our SEVC augments the base MVs and the spatial feature synergistically to generate hybrid contexts in feature space.
\section{Method}
\label{src:method}

\begin{figure}[tb]
	\centering
	\includegraphics[width=0.9\linewidth]{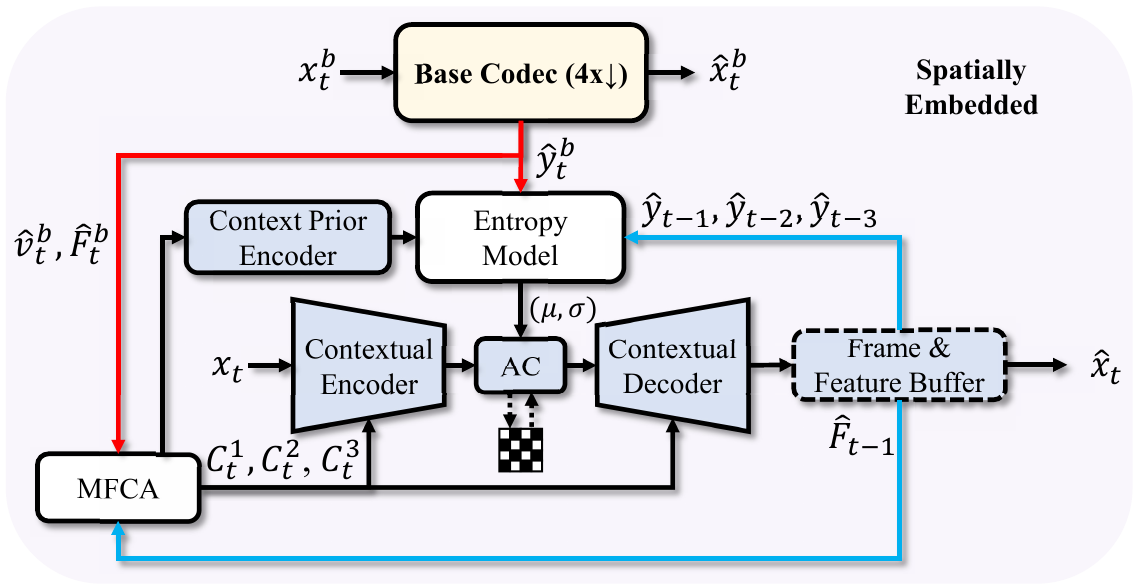}
	\vspace{-0.5em}
	\caption{
		Overview of our SEVC.
		Our SEVC embeds a base temporal-only codec for obtaining spatial references: base MVs $\hat{v}_t^b$, spatial feature $\hat{F}_t^b$, and spatial latent representation $\hat{y}_t^b$.
		Both the spatial and temporal references are utilized for the coding of the input frame $x_t$.
	}
	\vspace{-1.5em}
	\label{fig:fra}
\end{figure}

\subsection{Overview}
We downsample the input video $x$ to the low-resolution (LR) base video $x^b$, with their respective sources denoted as $X$ and $X^b$.
From an information-theoretic persepective~\cite{cover1999elements}, 
the entropy of $X$ can be decomposed into the sum of the entropy of $X^b$ and the conditional entropy of $X$ given $X^b$:
\begin{equation}
	\setlength{\abovedisplayskip}{3pt}
	\setlength{\belowdisplayskip}{3pt}
	H(X) = H(X^b) + H(X | X^b),
	\label{equ:1}
\end{equation}
which shows that the coding process for the original video can be divided into two parts:
one for the base video and another for the original video conditioned on the base video\footnote{The derivation of Equation~(\ref{equ:1}) is placed in the Supplementary Material.}.\par
Figure~\ref{fig:fra} shows the diagram of our spatially embedded video codec.
The base codec first compresses the 4x downsampling base frame $x_t^b$, deriving three spatial references:
the base MVs $\hat{v}_t^b$, the spatial feature $\hat{F}_t^b$, and the spatial latent representation $\hat{y}_t^b$.
These spatial references are subsequently augmented by the augmentative codec across two branches.
One is the progressive augmentation of $\hat{v}_t^b$ and $\hat{F}_t^b$ through our proposed MFCA module to generate hybrid spatial-temporal contexts $C_t^1, C_t^2, C_t^3$.
The other branch is the augmentation of $\hat{y}_t^b$ using multiple temporal latent representations $\hat{y}_{t-1},\hat{y}_{t-2},\hat{y}_{t-3}$ to generate the latent prior $\bar{y}_t$.
We will introduce them in Section~\ref{sec.MFCA} and Section~\ref{sec.lat}.
In addition, a joint spatial-temporal optimization strategy is proposed in Section~\ref{sec.two} to further boost RD performance.

\begin{figure*}[htb]
	\centering
	\includegraphics[width = 1.0\linewidth]{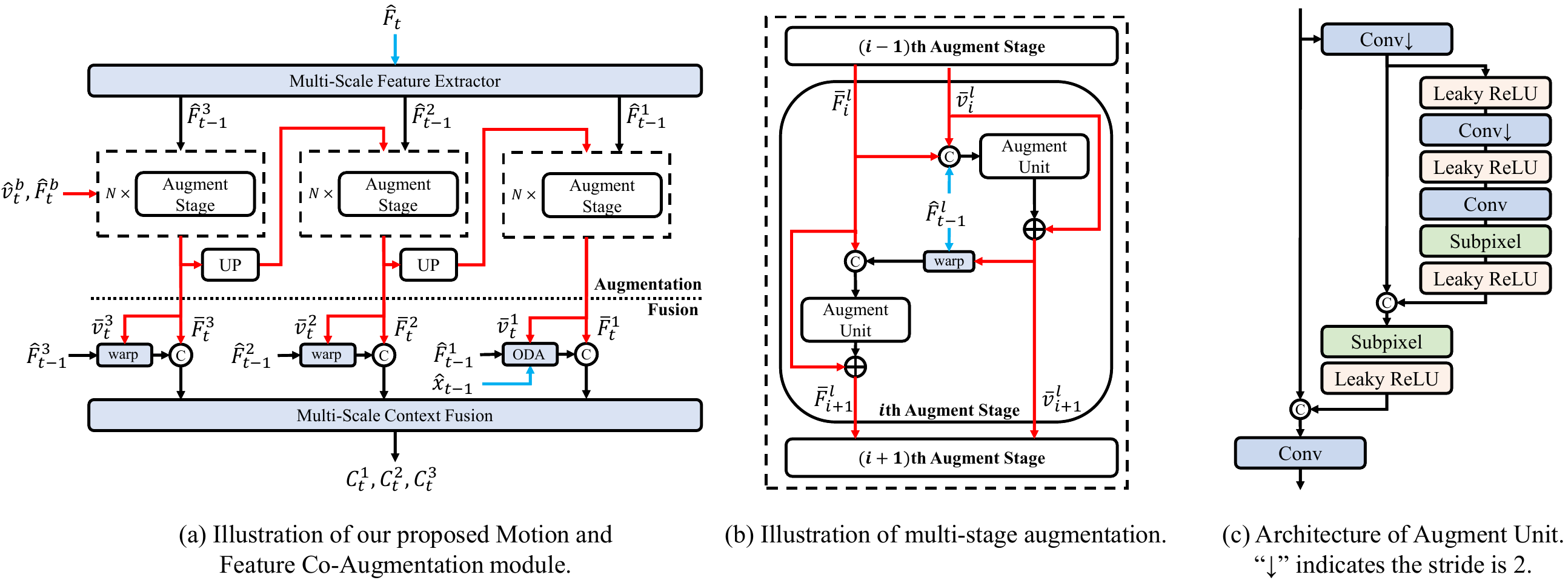}
	\vspace{-1.5em}
	\caption{Diagrams of the Motion and Feature Co-Augmentation module.
		``UP'' represents upsampling using subpixel convolution~\cite{shi2016real} with a factor of 2.
		``ODA'' represents Offset Diversity Alignment proposed in~\cite{dc}.
		\textcircled{\scriptsize{C}}~indicates channel dimension concatenation.
		\small{$\bigoplus$}~indicates element-wise addition.
	}
	\vspace{-1.5em}
	\label{fig.MFCA}
\end{figure*}

\begin{figure}[tb]
	\centering
	\includegraphics[width = 1.0\linewidth]{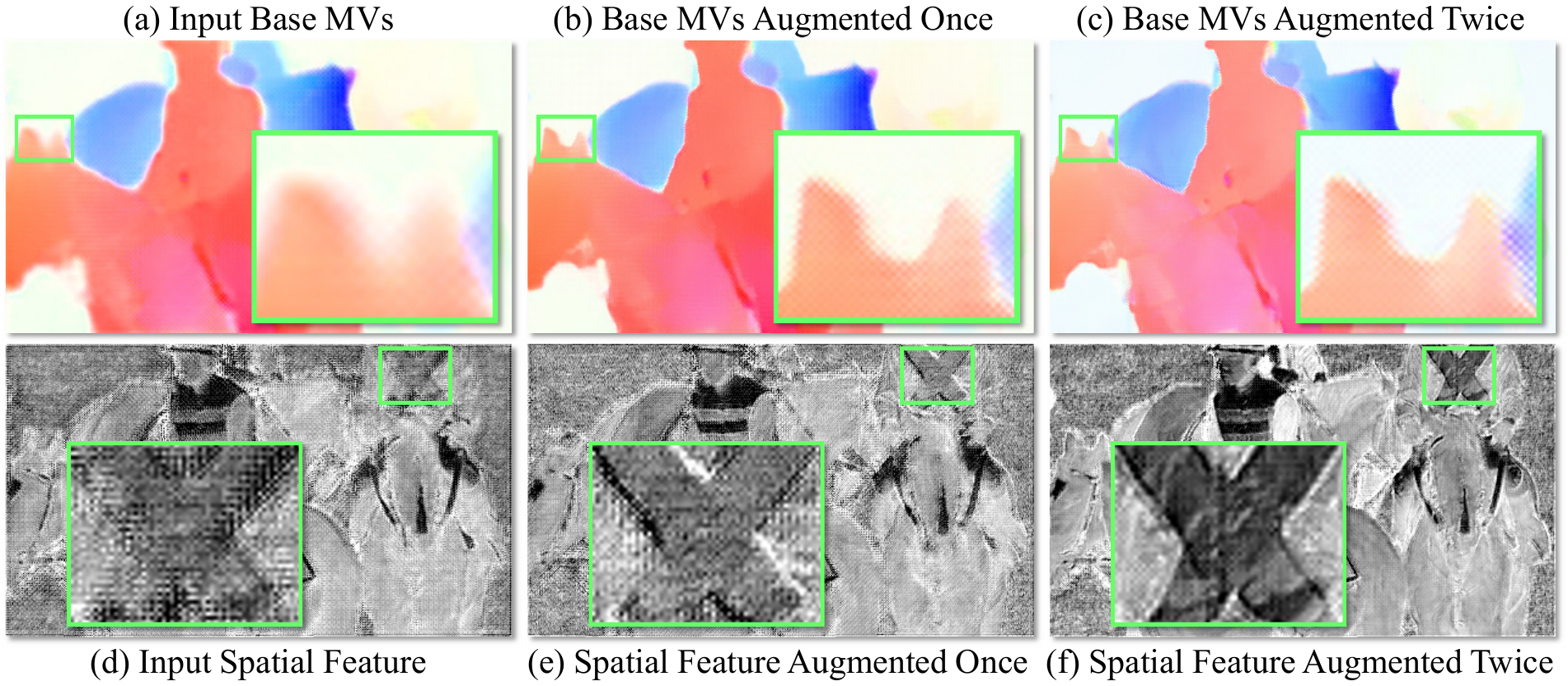}
	\vspace{-1.5em}
	\caption{
		Visualization of base MVs and the spatial feature within the multi-stage augmentation in the middle scale.
	}
	\vspace{-2.5em}
	\label{fig:vis_MFCA}
\end{figure}

\subsection{Motion and Feature Co-Augmentation\label{sec.MFCA}}
Most previous NVCs estimate and compress full-resolution MVs~\cite{dvc, dvc_pro, dcvc, tcm, hem,lin2020m,li2024neural,dc},
and reconstructed MVs are used for the alignment of decoded temporal frames or features.
However, these NVCs only depend on temporal references,
which can lead to suboptimal performance in complex scenarios involving large motions or emerging objects because of epistemic uncertainty~\cite{uncertainty}.
To alleviate this, \cite{fvc, hu2022coarse, dc} estimate offset map instead of optical flow to represent motion.
However, they still ignore motion coding errors.
Another aspect is that temporal-only references are not enough to describe emerging objects in the input frame that are not available in the previous frames.\par

Inspired by Video Super-Resolution~\cite{shi2022rethinking, Kappeler2016,chan2021basicvsr,chan2022basicvsr++},
We consider whether we can introduce LR base MVs and an LR spatial feature to provide basic information for reconstructing the input frame,
and then the temporal feature can be used to augment them progressively.
To this end, we propose the Motion and Feature Co-Augmentation (MFCA) module to progressively generate hybrid spatial-temporal deep contexts.

As shown in Figure~\ref{fig.MFCA}~(a), the MFCA module follows a bottom-up structure,
progressively augmenting the base MVs and the spatial feature with incremental qualities and resolutions.
The temporal feature $\hat{F}_{t-1}$ are first extracted into multi-scale features~\cite{tcm}.
Then starting from the smallest scale, the base MVs $\hat{v}_t^b$ and spatial feature $\hat{F}_t^b$ are augmented by the corresponding temporal feature through a multi-stage augmentation.
The multi-stage augmentation in each scale consists of several Augment Stages, and the operation within one Augment Stage is illustrated in Figure~\ref{fig.MFCA}~(b).
In the $ i$th Augment Stage of the $ l$th scale, the base MVs $\bar{v}_i^l$ are first augmented by adding the residual predicted through one Augment Unit:
\begin{equation}
	\setlength{\abovedisplayskip}{3pt}
	\setlength{\belowdisplayskip}{3pt}
	\bar{v}_{i+1}^l = \bar{v}_{i}^l + f_{AU}(cat(\hat{F}_{t-1}^l, \bar{F}_i^l, \bar{v}_i^l)), l = 1,2,3,
\end{equation}
where $cat$ represents channel dimension concatenation and $f_{AU}(\cdot)$ represents the Augment Unit.
The augmented base MVs $\bar{v}_{i+1}^l$ are then used for a better alignment of multi-scale features $\hat{F}_{t-1}^l$.
This better-aligned temporal feature could provide more accurate information to augment the spatial feature,
and the augmented spatial feature $\bar{F}_{i+1}^l$ is also obtained by adding the residual predicted through another Augment Unit:
\begin{equation}
	\setlength{\abovedisplayskip}{3pt}
	\setlength{\belowdisplayskip}{3pt}
	\bar{F}_{i+1}^l = \bar{F}_{i}^l + f_{AU}(cat(warp(\hat{F}_{t-1}^l, \bar{v}_{i+1}^l), \bar{F}_i^l)), l=1,2,3.
\end{equation}
The augmented base MVs $\bar{v}_{i+1}^l$ and spatial feature $\bar{F}_{i+1}^l$ are treated as low-quality inputs for the next Augment Stage and further augmented.
Since the resolutions differ across scales,
subpixel convolution~\cite{shi2016real} with a factor of 2 is used to ensure that the augmentation loop works properly.
The last augmented base MVs in each scale are utilized to align the temporal feature by warp operation or Offset Diversity Alignment~\cite{dc}.
The eventually aligned multi-scale temporal features are then concatenated with augmented spatial features,
fed into the context fusion module to generate hybrid spatial-temporal contexts $C_t^1, C_t^2, C_t^3$.
Although multi-stage augmentation may introduce considerable computational overhead, our Augment Unit,
as illustrated in Figure~\ref{fig.MFCA}~(c), operates at a declining resolution to mitigate this issue.\par

Figure~\ref{fig:vis_MFCA} presents a visualization of the multi-stage augmentation in the middle scale.
Before the augmentation, the base MVs and the spatial feature exhibited significant blurring at edges and regions with complex textures.
After two-stage augmentation, it can be observed that the blurring in the MVs is effectively mitigated.
The augmented MVs lead to a more accurate alignment of the temporal feature and thus a higher quality spatial feature.

\subsection{Spatial-Guided Latent Prior Augmentation\label{sec.lat}}
\begin{figure}[tb]
	\centering
	{\small
		\begin{minipage}[b]{4.5cm}
			\centering
			\centerline{\includegraphics[width = 1.0\linewidth]{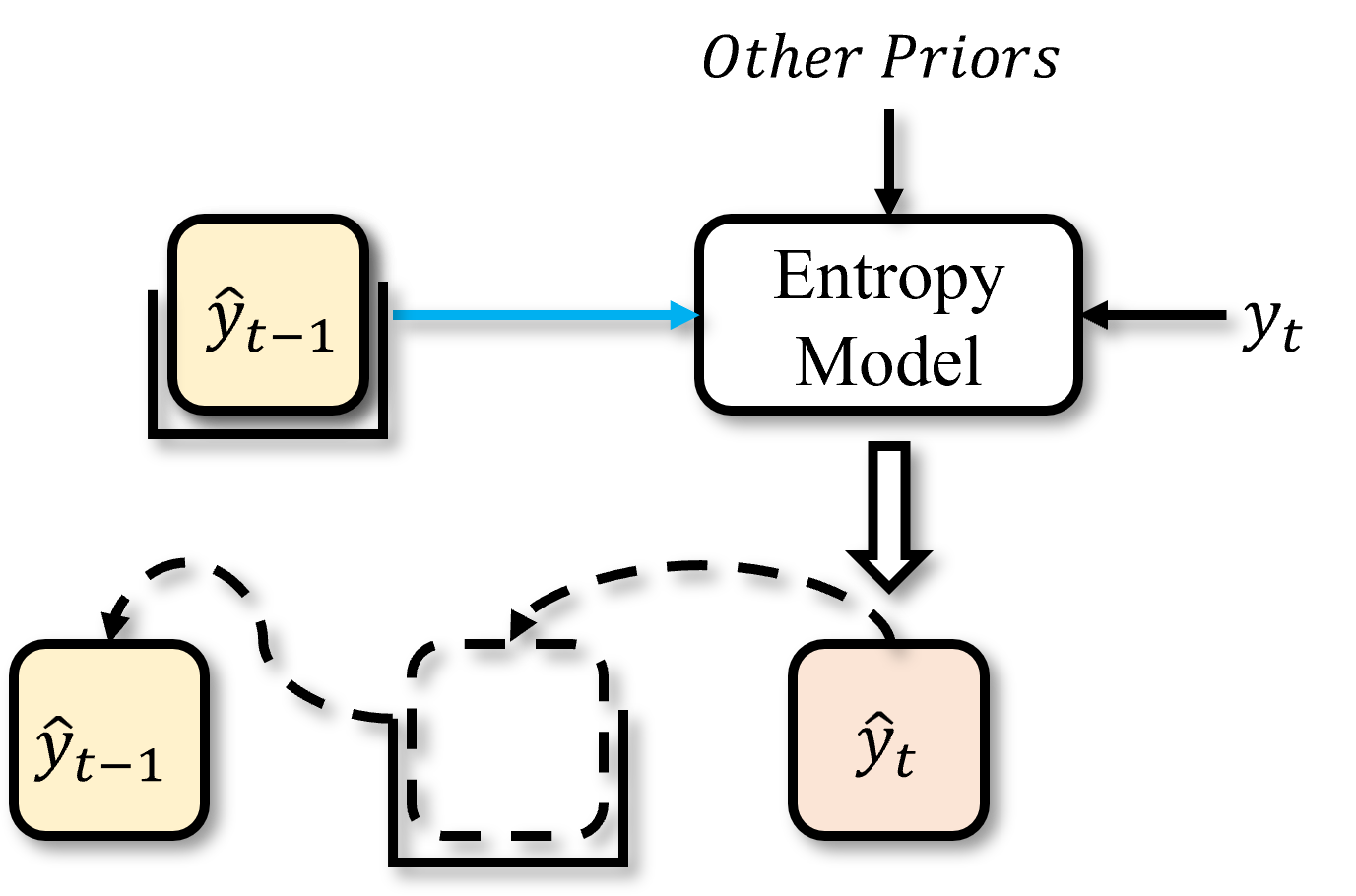}}
			\centerline{(a) Latent prior $\hat{y}_{t-1}$ used in previous NVCs.}\medskip
		\end{minipage} \\
		\begin{minipage}[b]{7.0cm}
			\centering
			\centerline{\includegraphics[width = 1.0\linewidth]{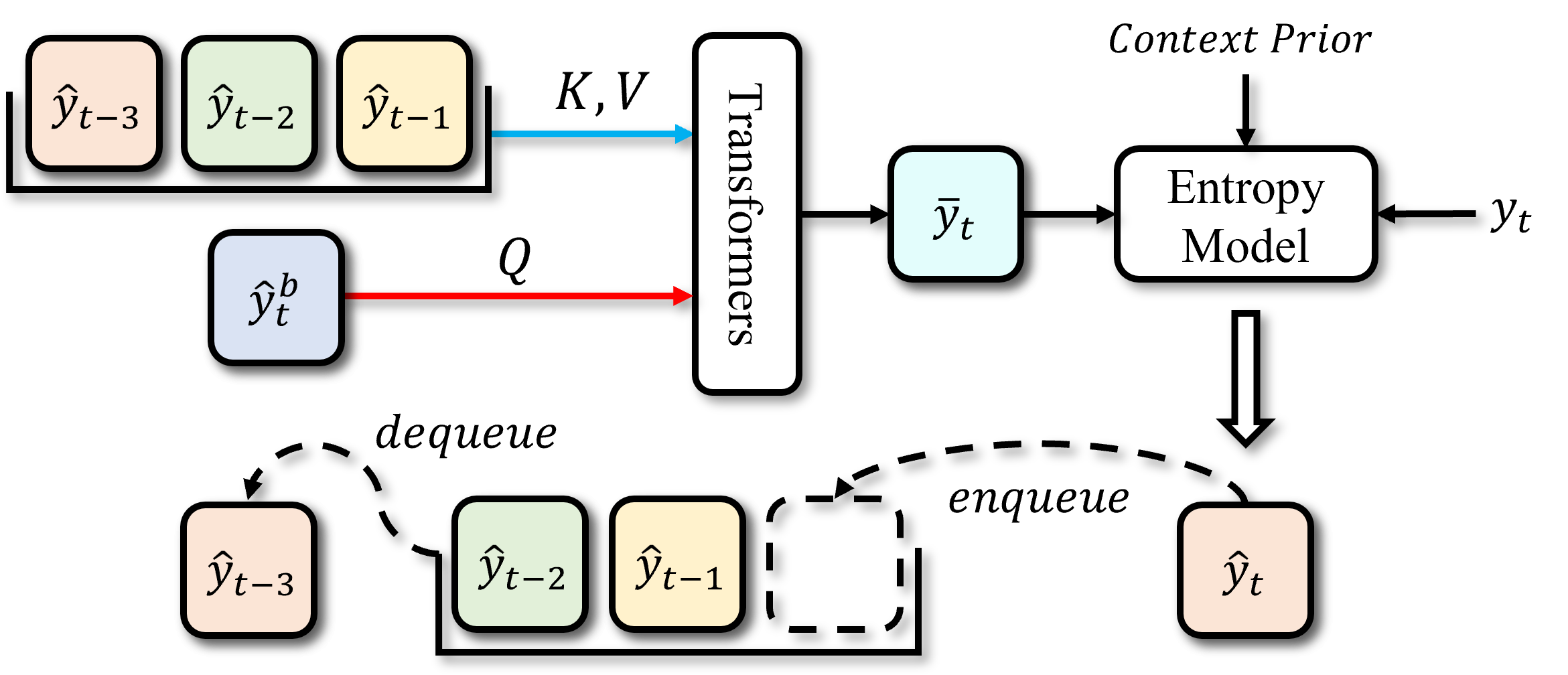}}
			\centerline{(b) Latent prior $\bar{y}_t$ used in our SEVC.}
		\end{minipage}
	}
	\vspace{-0.5em}
	\caption{
		Comparison of different latent priors.
		(a) is the temporally misaligned latent representation $\hat{y}_{t-1}$.
		Given the spatial latent representation $\hat{y}_t^b$ as the query,
		(b) is augmented by multiple temporal latent representations $\hat{y}_{t-1}, \hat{y}_{t-2}, \hat{y}_{t-3}$.
		Dashed lines indicate updates for the latent representations.
	}
	\vspace{-1.5em}
	\label{fig:latent}
\end{figure}

\begin{figure}[tb]
	\centering
	\includegraphics[width = 1.0 \linewidth]{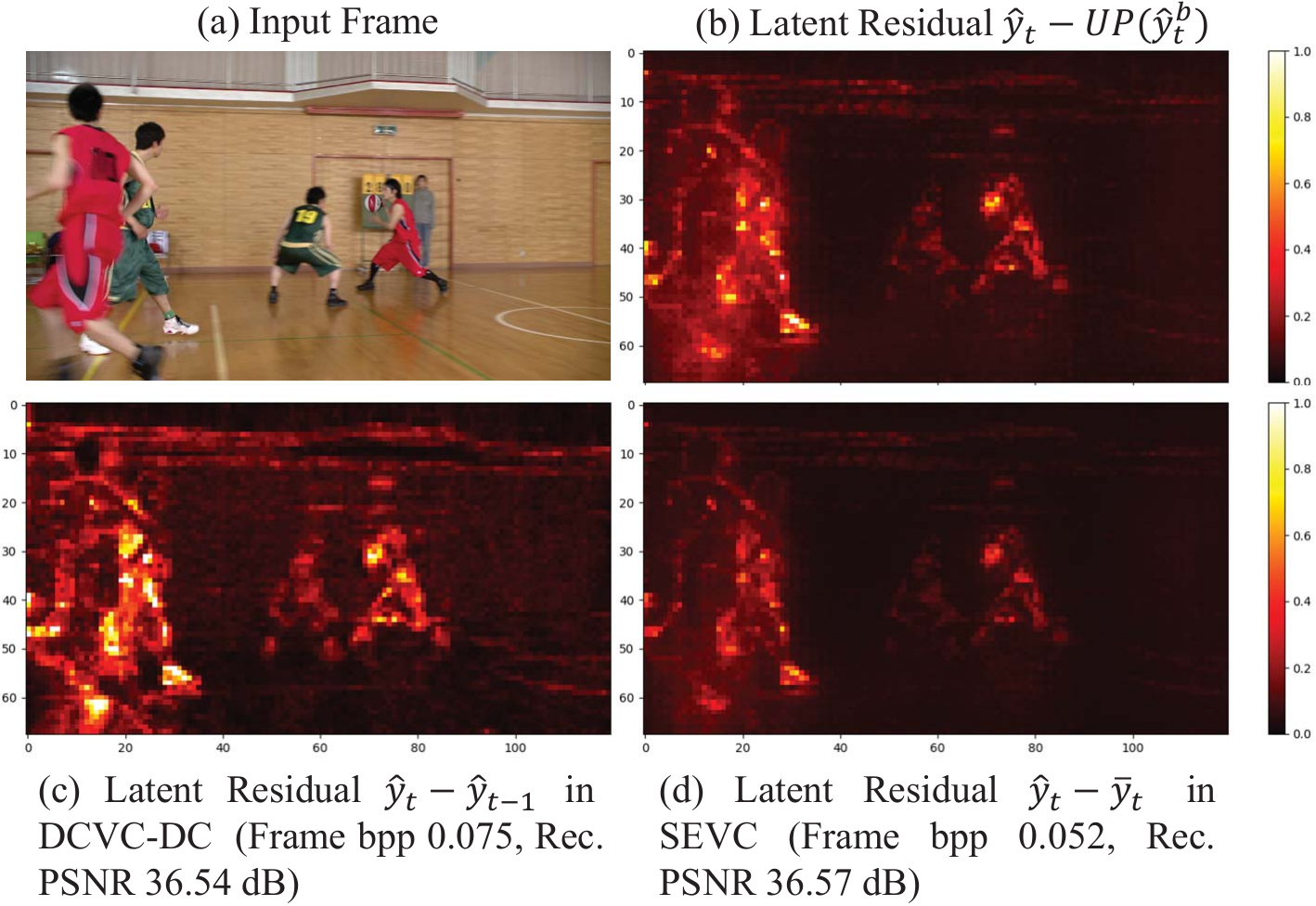}
	\vspace{-1.5em}
	\caption{
		Illustration of latent residuals for different priors.
		The residual of our latent prior $\bar{y}_t$ and latent representation $\hat{y}_t$ is much less than
		that in DCVC-DC.
		Compared to the upsampled spatial latent representation $UP(\hat{y}_t^b)$, our prior is augmented by temporal latent representations and also performs a less residual.
	}
	\vspace{-1.5em}
	\label{fig:latent_residual}
\end{figure}

Entropy models in NVCs rely on the prior information to estimate the distribution of the quantized latent representation $\hat{y}_t$.
Subsequently, an entropy coding algorithm~\cite{ans} can be used to compress $\hat{y}_t$ into a bitstream.
The accuracy and richness of the priors significantly affect the precision of the estimated distribution.
In addition to the commonly used context prior~\cite{dcvc} and the hyperprior~\cite{balle2018variational},
as shown in Figure~\ref{fig:latent} (a), previous NVCs~\cite{hem,dc,li2024neural,ssd} consider the previously decoded latent representation $\hat{y}_{t-1}$ as the latent prior.
This off-the-shelf latent prior is not optimal because of the misalignment with the current index.
As show in Figure~\ref{fig:latent_residual} (c), a large residual occurs between $\hat{y}_t$ and the latent prior $\hat{y}_{t-1}$,
adversely affecting distribution estimation.\par

In our SEVC, in addition to the temporal latent representation $\hat{y}_{t-1}$, there is a spatial latent representation $\hat{y}_t^b$.
Shi \textit{et al.}~\cite{shi2022rethinking} gives a perspective that the attention mechanism in Transformers~\cite{vaswani2017attention,vit,liu2021swin}
can directly capture the temporal correlation between multiple frames.
Unlike existing NVCs, which lack a suitable low-quality representation of the input frame for querying,
our spatial latent representation $y_t^b$ fulfills this requirement well.\par

The spatial latent representation $\hat{y}_t^b$ not only guides the alignment of the temporal latent representation $\hat{y}_t$,
but also could complement richer prior information.
To further extend prior information, we also query for multiple temporal latent representations.
As shown in Figure~\ref{fig:latent}~(b), given $\hat{y}_t^b$ (4x upsampled by subpixel convolution to align the resolution) as the query $Q$
and three temporal latent representations $\hat{y}_{t-3}, \hat{y}_{t-2}, \hat{y}_{t-1}$ as the keys $K$ and values $V$,
the Transformers~\cite{shi2022rethinking,liang2021swinir} align and fuse multiple temporal latent representations as the latent prior $\bar{y}_t$.
The temporal latent representations are organized in a queue structure.\par

As shown in Figure~\ref{fig:latent_residual}, our latent prior $\bar{y}_t$ has a less residual compared to the latent prior $\hat{y}_{t-1}$ used in previous NVCs and the upsampled spatial latent representation $UP(\hat{y}_t^b$).
The latent prior and context prior extracted from hybrid contexts, will be fed together into the entropy model to estimate the distribution.
Due to similar characteristics of hyper encoder/decoder and our base codec,
we discard the hyperprior to avoid posterior collapse~\cite{lucas2019don}.

\begin{figure*}[tb]
	\centering
	\includegraphics[width = 1.0\linewidth]{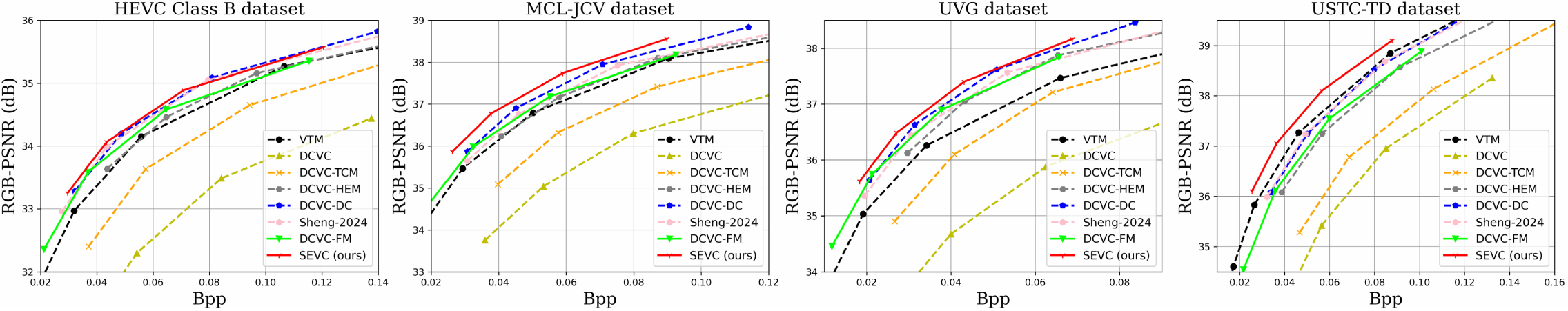}
	\vspace{-2.0em}
	\caption{Rate and distortion curves on four 1080p datasets. The Intra Period is 32 with 96 frames.
	}
	\vspace{-0.5em}
	\label{fig.RD1}
\end{figure*}

\begin{table*}[tb]
	\centering
	\caption{BD-Rate (\%) comparison for PSNR. The Intra Period is 32 with 96 frames. The anchor is VTM-13.2 LDB.\label{tab.bd1}}
	\vspace{-0.5em}
	\begin{threeparttable} 
		{\small
			\begin{tabular}{lccccccccc}
				\toprule[1.0pt]
				                                   & HEVC~B      & HEVC~C      & HEVC~D      & HEVC~E      & MCL-JCV     & UVG         & USTC-TD     & Average     \\ \midrule
				\small DCVC~\cite{dcvc}            & 119.6       & 152.5       & 110.9       & 274.8       & 106.6       & 133.9       & 139.6       & 148.3       \\ \midrule
				\small DCVC-TCM~\cite{tcm}         & 32.8        & 62.1        & 29.0        & 75.7        & 38.2        & 23.1        & 75.3        & 48.0        \\ \midrule
				\small DCVC-HEM~\cite{hem}         & --0.7       & 16.1        & --7.1       & 20.7        & --1.6       & --17.2      & 20.6        & 4.4         \\ \midrule
				\small DCVC-DC~\cite{dc}           & --13.9      & --8.8       & --27.7      & --19.2      & --14.4      & --25.9      & 10.8         & --14.2      \\ \midrule
				\small Sheng-2024~\cite{ssd}       & --13.7      & --2.3       & --24.9      & --8.4       & --7.1       & --19.7      & 7.7         & --9.8       \\ \midrule
				\small DCVC-FM$^{\dag}$~\cite{li2024neural} & --10.3      & --8.4       & --25.8      & --21.9      & --8.1       & --20.4      & 25.7        & --9.9      \\ \midrule
				\small SEVC (ours)                 & \bf{--16.4} & \bf{--15.8} & \bf{--30.0} & \bf{--28.5} & \bf{--23.3} & \bf{--30.2} & \bf{--13.4} & \bf{--22.5} \\
				\bottomrule[1.0pt]
			\end{tabular}
		}
		\begin{tablenotes}
			\scriptsize
			\item[\dag] Since the YUV MSE weight is higher than the RGB MSE in the loss function of DCVC-FM, its performance is inferior to DCVC-DC in this setting.
		\end{tablenotes}
	\end{threeparttable}
	\vspace{-1.5em}
\end{table*}

\begin{figure}[tb]
	\centering
	\includegraphics[width = 1.0 \linewidth]{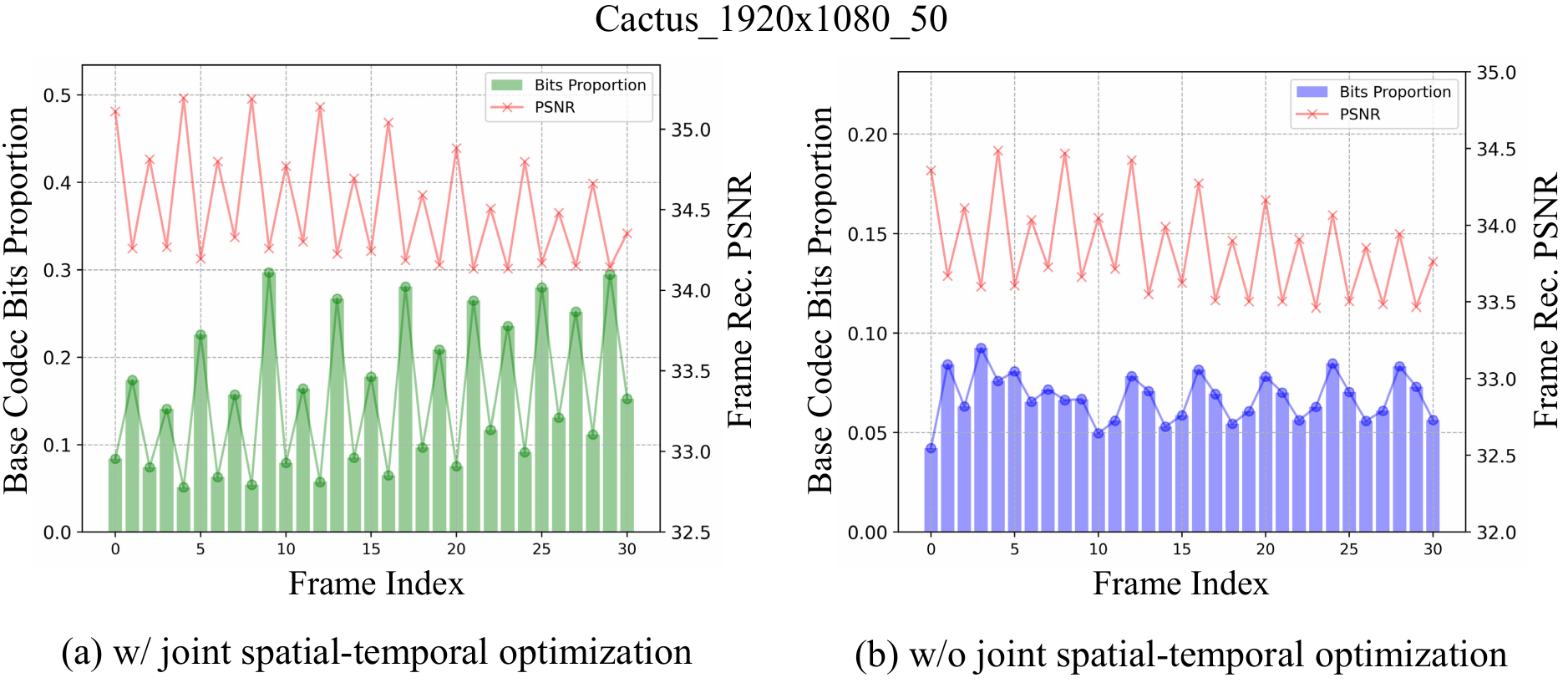}
	\vspace{-1.5em}
	\caption{
		Illustration of base codec bits proportion.
		The joint spatial-temporal optimization leads our codec to learn a quality-adaptive bits allocation.
		Given the temporally hierarchical quality structure,
		the low-quality reconstruction frame implies more base bits while the high-quality one favors less.
	}
	\vspace{-1.5em}
	\label{fig:two-layer}
\end{figure}

\begin{figure*}[tb]
	\centering
	\includegraphics[width = 1.0\linewidth]{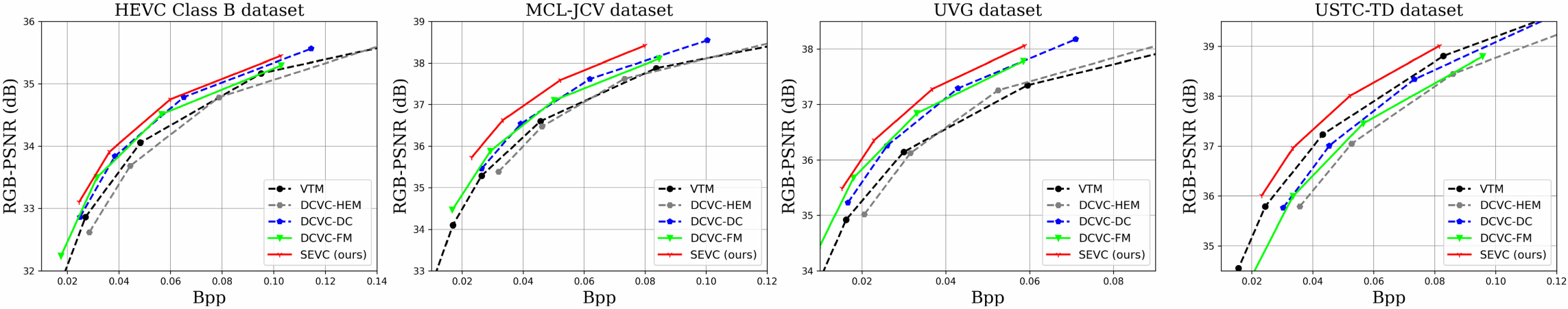}
	\vspace{-2.0em}
	\caption{Rate and distortion curves on four 1080p datasets. The Intra Period is --1 with 96 frames.
	}
	\vspace{-0.5em}
	\label{fig.RD2}
\end{figure*}

\begin{table*}[tb]
	\centering
	\caption{BD-Rate (\%) comparison for PSNR. The Intra Period is --1 with 96 frames. The anchor is VTM-13.2 LDB.\label{tab.bd2}}
	\vspace{-0.5em}
	{\small
		\begin{tabular}{lcccccccc}
			\toprule[1.0pt]
			                                   & HEVC~B      & HEVC~C      & HEVC~D      & HEVC~E      & MCL-JCV     & UVG         & USTC-TD     & Average     \\ \midrule
			\small DCVC-HEM~\cite{hem}         & 10.0        & 30.0        & --1.1       & 68.6        & 4.9         & 1.2         & 27.2        & 20.1        \\ \midrule
			\small DCVC-DC~\cite{dc}           & --10.8      & --0.1       & --24.2      & --7.7       & --13.0      & --21.2      & 11.9        & --9.3       \\ \midrule
			\small DCVC-FM~\cite{li2024neural} & --11.7      & --8.2       & --28.5      & --26.6      & --12.5      & --24.3      & 23.9        & --12.6      \\ \midrule
			\small SEVC (ours)                 & \bf{--17.5} & \bf{--15.1} & \bf{--31.6} & \bf{--34.0} & \bf{--27.7} & \bf{--33.2} & \bf{--12.5} & \bf{--24.5} \\
			\bottomrule[1.0pt]
		\end{tabular}
	}
	\vspace{-1.5em}
\end{table*}

\begin{table}[ht]
	\centering
	\caption{BD-Rate (\%) comparison for PSNR. The Intra Period is 32 with 96 frames. The anchor is DCVC-HEM.\label{tab.part_seq}}
	\vspace{-0.5em}
	\begin{threeparttable}
		{\footnotesize
			\begin{tabular}{lcc}
				\toprule[1.0pt]
				Sequence Name                                                    & \hspace{-1em} DCVC-DC & SEVC   \\ \midrule
				*~\textit{USTC\_BicycleDriving}~\cite{li2024ustc}                & \hspace{-1em} --17.9  & --55.9 \\ \midrule
				*~\textit{videoSRC21\_1920x1080\_24}~\cite{wang2016mcl}          & \hspace{-1em} --14.5  & --31.3 \\ \midrule
				\dag~\textit{BasketballDrive\_1920x1080}\_50~\cite{hevc_dataset} & \hspace{-1em} --8.5   & --24.7 \\ \midrule
				\dag~\textit{BQMall\_832x480\_60}~\cite{hevc_dataset}            & \hspace{-1em} --28.7  & --35.9 \\
				\bottomrule[1.0pt]
			\end{tabular}
		}
		\begin{tablenotes}
			\scriptsize
			\item[*] indicates sequences including large motions.
			\item[\dag] indicates sequences including significant emerging objects.
		\end{tablenotes}
	\end{threeparttable}
	\vspace{-1.5em}
\end{table}

\subsection{Joint Spatial-Temporal Optimization\label{sec.two}}
In traditional spatially scalable video coding~\cite{vvc_spatial,shvc,svc},
when videos with two different resolutions are compressed,
the LR video is commonly given a higher quality to provide better spatial references.
Better spatial references benefit the coding for the full-resolution video and improve the performance.
Such a hierarchical quality structure is usually accomplished by assigning a smaller Quantization Parameter (QP) to the base layer,
thereby allocating more bits to it~\cite{sullivan2013standardized}.\par

However, our SEVC differs from those traditional codecs in two aspects.
One is that the performance of the base codec can be sacrificed for the purpose of obtaining a better full-resolution RD performance.
Another aspect is that the bit allocation strategy for spatial references could be optimized end-to-end without handcrafted QP design.\par

To stabilize the training, we first independently optimize the two parts like previous NVCs~\cite{dvc, dvc_pro, dcvc, tcm, ho2022canf, fvc,hu2022coarse}.
\begin{equation}
	\setlength{\abovedisplayskip}{4pt}
	\setlength{\belowdisplayskip}{4pt}
	L = \left\{
	\begin{aligned}
		\frac{1}{T} \sum_{t}^{T} (w_t \cdot \lambda \cdot D_t^b + R_t^b), & ~~\text{if base codec}, \\
		\frac{1}{T} \sum_{t}^{T} (w_t \cdot \lambda \cdot D_t + R_t),     & ~~\text{otherwise},
	\end{aligned}
	\label{equ:indepent}
	\right.
\end{equation}
where $w_t$ denotes the frame-level hierarchical quality weight~\cite{dc} and $t$ indicates the frame index.
$R_t^b$ is the estimated bpp (bits per pixel) for the base codec and $R_t$ for the whole codec.
$D_t^b$ measures the distortion between LR input frame $x_t^b$ and LR reconstructed frame $\hat{x}_t^b$,
while $D_t$ measures it between input frame $x_t$ and reconstructed frame $\hat{x}_t$.
The base codec is trained first, followed by the augmentative part.
When one part is being trained, the parameters of the other are frozen.
After the independent optimization, we conduct the following joint spatial-temporal optimization:
\begin{equation}
	\setlength{\abovedisplayskip}{3pt}
	\setlength{\belowdisplayskip}{3pt}
	L = \frac{1}{T} \sum_{t}^{T}(w_t \cdot \lambda \cdot (D_t + w_l \cdot D_t^b) + R_t),
	\label{equ:joint}
\end{equation}
where $w_l$ is a regular constraint of the base reconstruction.
$w_l$ is set to a small value to ensure the base reconstruction while barely affecting the full-resolution RD performance.
Experiments in Sec.~\ref{ab.3} show that this constraint is necessary to guarantee proper spatial references.\par

As shown in Figure~\ref {fig:two-layer}, without joint optimization, the bits proportion shows a relatively flat trend.
However, after conducting joint optimization, our SEVC learned a more reasonable bit allocation.
Given the temporally hierarchical quality structure,
the low-quality reconstruction implies more base bits while the high-quality one favors less.
This suggests that our codec learned to adaptively allocate bits for spatial references based on the required quality.

\begin{figure*}[tb]
	\centering
	\includegraphics[width = 1.0 \linewidth]{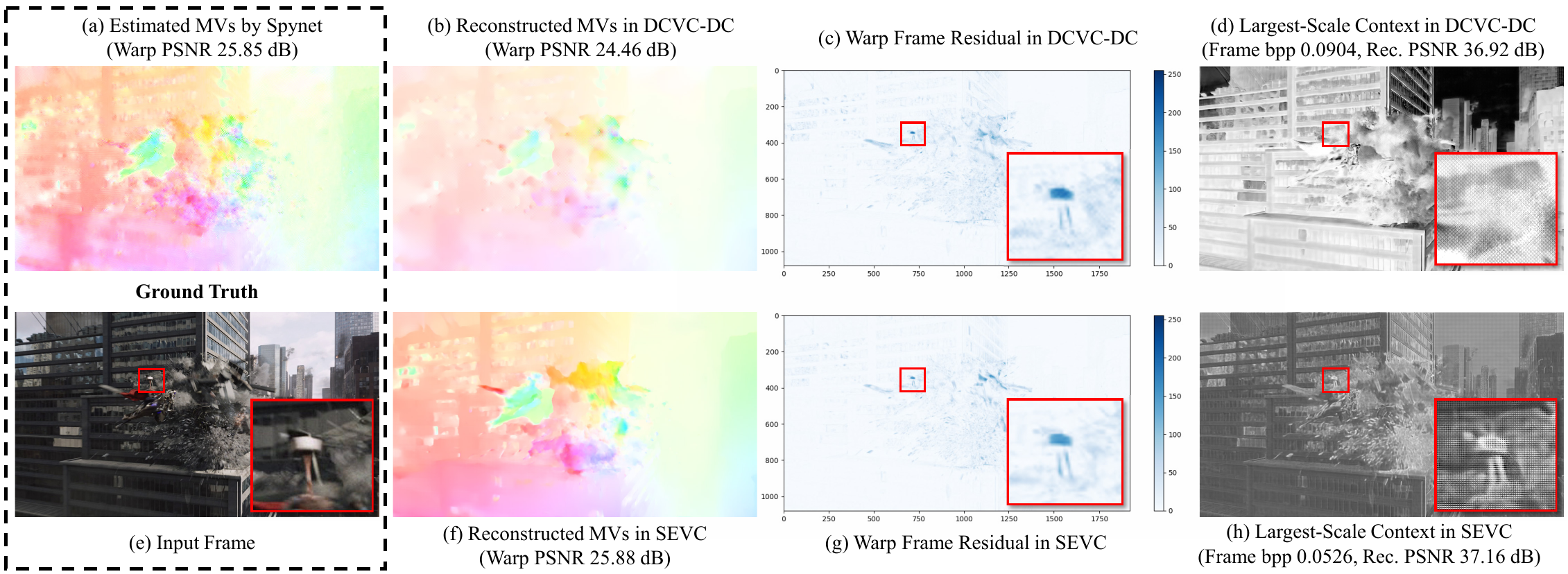}
	\vspace{-2.0em}
	\caption{
		Visualization of the MVs and contexts in DCVC-DC and our SEVC.
	}
	\vspace{-1.5em}
	\label{fig:com_dc}
\end{figure*}
\section{Experimental Results}
\label{sec:experiment}
\subsection{Experimental Settings}
\textbf{Training Setup}
Our proposed SEVC is trained on Vimeo-90k~\cite{xue2019video} using the multi-stage training strategy~\cite{tcm}.
The independent optimization for the two parts is conducted on the official train split, 
while the joint optimization is conducted on a selected subset of 9000 sequences from the original videos of the Vimeo-90k dataset.
The sequences are randomly cropped into 256$\times$256 in the independent optimization training stage and 384$\times$256 in the joint optimization finetuning stage.
A group of 6 frames is used for independent optimization and 32 frames for joint optimization.
The $\lambda$ in Equation~(\ref{equ:indepent}) and (\ref{equ:joint}) is set to \{50, 95, 200, 400\} for different bitrates, and the batch size is set to 8.
The LR input frame is generated by bicubic downsampling~\cite{cubic}.

\textbf{Testing Setup}
UVG~\cite{mercat2020uvg}, MCL-JCV~\cite{wang2016mcl}, HEVC Class B$\sim$E~\cite{hevc_dataset}, and USTC-TD~\cite{li2024ustc} are used as the test sets.
The distortion is calculated on full-resolution RGB colorspace and the BT.601 coefficient is used for the conversion between YUV420 and RGB colorspace.
To demonstrate the superiority of our proposed SEVC,
we compare our codec with the traditional codec--VTM-13.2 LDB~\cite{VTM} and
several advanced conditional NVCs designed for low-delay scenarios: DCVC~\cite{dcvc}, DCVC-TCM~\cite{tcm}, DCVC-HEM~\cite{hem}, DCVC-DC~\cite{dc}, Sheng-2024~\cite{ssd}, and DCVC-FM~\cite{li2024neural}.
We follow~\cite{tcm, hem, ssd, dc,li2024neural} to test 96 frames for each video.
The Intra Period (IP) is set to 32 when SEVC is compared to all codecs and IP is set to --1 when compared to DCVC-DC and DCVC-FM.

\subsection{Compared to Previous NVCs}
Figure~\ref{fig.RD1} and Figure~\ref{fig.RD2} shows the RD curves of our SEVC and other codecs on four 1080p datasets,
and Table~\ref{tab.bd1} and Table~\ref{tab.bd2} shows the BD-Rate results on all datasets.
When the IP is set to 32, DCVC-DC achieves the highest bitrate saving compared to other previous NVCs, averaging 14.2\% over the anchor VTM.
However, our SEVC achieves 8.3\% more bitrate saving compared to DCVC-DC.
Since DCVC-FM is mainly designed for IP --1,
DCVC-FM achieves the highest bitrate saving under IP --1, averaging 12.6\% over the anchor VTM.
In this case, our SEVC still achieves 11.9\% more bitrate saving.
In addition to the superior performance, our SEVC provides an additional base bitstream which can be independently decoded into a low-resolution video.\par

Furthermore, as demonstrated in Table~\ref{tab.part_seq},
our SEVC far surpasses the previous SOTA codec in sequences with large motions or significant emerging objects.
Through the visualization of intermediate MVs and contexts in Figure~\ref{fig:com_dc},
we can observe that our augmented MVs have an even higher quality compared to MVs estimated by Spynet~\cite{ranjan2017optical}.
Although better MVs were obtained, certain areas that newly appeared, such as the Thor hammer marked by the red box, still exhibit a large residual.
This indicates that temporal references are not rich enough to generate a good prediction for the emerging object.
However, our hybrid spatial-temporal contexts incorporate spatial references and complement the additional description for the hammer.
More visualization results can be found in the Supplementary Material.

\subsection{Ablation Studies}
To validate the effectiveness of each technique used to improve the performance of our SEVC, we conduct comprehensive ablation studies.
The average BD-Rate, calculated in terms of RGB PSNR on the HEVC datasets, is used here for comparison.
Our method is marked in bold.\par

\textbf{Motion and Feature Co-Augmentation.\label{ab.1}}
Table~\ref{tab.ab1} shows the ablation study on the effectiveness of our MFCA.
``w/o augmentation'' represents that the base MVs and the spatial feature are directly sent to the context fusion part without any augmentation.
``$N$ stages'' represents $N$ Augment Stages in each scale.
A positive correlation can be observed between performance and the number of Augment Stages.
However, compromising complexity and performance, we adopt two Augment Stages in each scale.

\textbf{Different Latent Priors.\label{ab.2}}
To demonstrate the advantages of our spatial-guided latent prior,
we split and reassemble the prior generation components into four methods in Table~\ref{tab.ab2}.
The comparison between $M_1$ and $M_2$ indicates benefits brought by multiple temporal latent representations.
$M_3$ uses convolution networks instead of Transformers in $M_2$, resulting in a 5\% increase in bitrate, which demonstrates the effectiveness of
implicit alignment of Transformers.
The last $M_4$ method further discards spatial latent representation in prior generation, and also performs bitrate increase.\par

\begin{table}[tb]
	\centering
	\caption{Ablation Study on MFCA.\label{tab.ab1}}
	\vspace{-0.5em}
	{\scriptsize
		\begin{tabular}{lcccc}
			\toprule[1.0pt]
			             & w/o augmentation & 1 stages & \textbf{2 stages} & 3 stages \\ \midrule
			MACs         & 2616G      & 2938G    & \textbf{3263G}    & 3588G    \\ \midrule
			BD-Rate (\%) & 17.5       & 4.7      & \textbf{0.0}      & --1.9    \\
			\bottomrule[1.0pt]
		\end{tabular}
	}
\end{table}

\begin{table}[tb]
	\centering
	\caption{Ablation Study on Different Latent Priors.\label{tab.ab2}}
	\vspace{-0.5em}
	{\scriptsize
		\begin{tabular}{lcccc}
			\toprule[1.0pt]
			                                                    & $\bm{M_1}$   & $M_2$       & $M_3$       & $M_4$       \\ \midrule
			$\hat{y}_t^b$                                       & \(\bullet\)  & \(\bullet\) & \(\bullet\) &             \\ \midrule
			$\hat{y}_{t-1}$~\cite{hem}                        &              & \(\bullet\) & \(\bullet\) & \(\bullet\) \\ \midrule
			$\hat{y}_{t-1}, \hat{y}_{t-2}, \hat{y}_{t-3}$ & \(\bullet\)  &             &             &             \\ \midrule
			Transformers~\cite{shi2022rethinking}               & \(\bullet\)  & \(\bullet\) &             &             \\ \midrule
			BD-Rate (\%)                                        & \textbf{0.0} & 1.5         & 6.5         & 8.4         \\
			\bottomrule[1.0pt]
		\end{tabular}
	}
	\vspace{-1.5em}
\end{table}

\textbf{Joint Spatial-Temporal Optimization.\label{ab.3}}
To better understand how the joint spatial-temporal optimization affects the RD performance, 
we conduct experiments under different strategies as shown in Table~\ref{tab.ab3}.
If the joint optimization is not conducted, i.e., the base codec parameter is fixed during finetuning,
the base codec occupies 13\% bitrate.
This is almost the same as setting $w_l$ in Equation~(\ref{equ:joint}) to 0.01.
Furthermore, a modest increment in $w_l$ to 0.05 and 0.1 permits additional bits allocated to the base codec, 
facilitating a hierarchical quality structure akin to that achieved in traditional spatially scalable video coding.
However, unconstrained setting with $w_l=0$ or excessive bits allocated to the base codec has detrimental effects.

\subsection{Complexity}
Table~\ref{tab.complexity} shows the MACs, encoding time (ET), and decoding time (DT) comparison.
The MACs of our SEVC are about the same as DCVC-HEM but reduce 44.6\% more bitrate compared to VTM.
Although compressing two videos, our MACs are still reduced by 15.4\% compared to Sheng-2024~\cite{ssd}.
However, our SEVC is inferior to DCVC-DC and DCVC-FM, which made excellent optimizations for complexity.
Considering the higher compression ratio and additional functionality for fast decoding to quickly skim through a video, the increase in complexity is worthwhile.

\begin{table}[tb]
	\centering
	\caption{Ablation Study on Joint Spatial-Temporal Optimization.\label{tab.ab3}}
	\vspace{-0.5em}
	{\scriptsize
	\renewcommand{\arraystretch}{1.2}
		\begin{tabular}{lcccccc}
			\toprule[1.0pt]
			             & w/o joint           &                      &                        & $w_l$                         &                       &                      \\
						 \cline{3-7}
			             & optimization        & \hspace{-0.5em} $0$  & \hspace{-0.5em} $0.01$ & $\bm{0.05}$                   & \hspace{-0.5em} $0.1$ & \hspace{-0.5em}$1.0$ \\ \midrule
			\makecell{
			Base bitrate                                                                                                                                                        \\ proportion
			}            & \hspace{-0.5em}13\% & \hspace{-0.5em} 8\%  & 12 \%                  & \hspace{-0.5em} \textbf{23\%} & 31\%                  & 54\%                 \\ \midrule
			BD-Rate (\%) & \hspace{-0.5em} 5.1 & \hspace{-0.5em} 11.6 & \hspace{-0.5em} 5.3    & \hspace{-0.5em} \textbf{0.0}  & \hspace{-0.5em} 1.2   & \hspace{-0.5em}5.0   \\
			\bottomrule[1.0pt]
		\end{tabular}
	}
\end{table}

\begin{table}[tb]
	\centering
	\caption{Complexity Comparison.\label{tab.complexity}}
	\vspace{-0.5em}
	\begin{threeparttable}
		{\scriptsize
			\begin{tabular}{lcccccc}
				\toprule[1.0pt]
				     & \hspace{-0.5em}DCVC-HEM & \hspace{-0.5em}DCVC-DC & \hspace{-0.5em} Sheng-2024 & \hspace{-0.5em}DCVC-FM & \hspace{-0.5em}\textbf{SEVC}  \\ \midrule
				MACs & \hspace{-0.5em}3435G    & \hspace{-0.5em}2764G   & \hspace{-0.5em}3830G       & \hspace{-0.5em} 2334G  & \hspace{-0.5em}\textbf{3263G} \\\midrule
				ET   & \hspace{-0.5em} 548ms   & \hspace{-0.5em}663ms   & \hspace{-0.5em}968ms       & \hspace{-0.5em} 587ms  & \hspace{-0.5em}\textbf{775ms} \\\midrule
				DT   & \hspace{-0.5em} 213ms   & \hspace{-0.5em}557ms   & \hspace{-0.5em}775ms       & \hspace{-0.5em} 495ms  & \hspace{-0.5em}\textbf{734ms} \\
				\bottomrule[1.0pt]
			\end{tabular}
			\begin{tablenotes}
				\item[1] Tested on a NVIDA RTX 3090 GPU with 1080p sequences as inputs.
			\end{tablenotes}
		}
	\end{threeparttable}
	\vspace{-1.5em}
\end{table}

\section{Conclusion}
\label{src:conclusion}
In this paper, we present our solution to augmenting deep contexts and the latent prior for frame coding. 
For augmenting contexts, we synergistically augment the MVs and the spatial feature to generate hybrid spatial-temporal contexts. 
For augmenting the latent prior, we employ the spatial latent representation to merge multiple temporal latent representations.
Finally, a joint spatial-temporal optimization is conducted to adjust the bit allocation for spatial references.
Our SEVC well alleviates the limitations in handling large motions or emerging objects, 
achieving an 11.9\% more bitrate saving compared to the previous SOTA codec.\par

However, the base codec used in our SEVC is limited to an existing NVC.
Although spatial references can be obtained directly from the reconstruction of low-resolution video,
the reconstruction limits the characteristics of spatial references.
In the future, we will investigate more efficient ways to model and extract spatial references.\par
{
    \small
    \bibliographystyle{ieeenat_fullname}
    \bibliography{refs}
}

\newpage
\section*{\Large Supplementary Material}
\renewcommand{\thesection}{\Alph{section}} 
\renewcommand{\theequation}{\alph{equation}} 
\renewcommand{\thefigure}{\alph{figure}} 
\renewcommand{\thetable}{\alph{table}} 

\setcounter{equation}{0}
\setcounter{figure}{0}
\setcounter{table}{0}
\setcounter{section}{0}
This supplementary material document provides additional
details of our proposed Spatially Embedded Video Codec (SEVC).
The remainder of the supplementary material
is divided into three parts.
Section~\ref{sec:test} gives the configuration of the traditional codec---VTM-13.2~\cite{VTM}.
Section~\ref{sec:net} gives the detailed network architectures of our proposed modules.
Section~\ref{sec:derivation} gives the derivation of Euqation~(\textcolor{red}{1}).
Section~\ref{sec:add_result} provides additional comparison results.

\section{Configuration of the Traditional Codec\label{sec:test}}
When testing traditional codec---VTM-13.2~\cite{VTM}, the input video sequences are in YUV444 format to achieve a better compression ratio~\cite{tcm,hem,dc}.
The YUV444 video sequences are converted from the RGB video sequences, which are used as the inputs of NVCs.
The configuration parameters for encoding each video are as:
	\begin{itemize}[leftmargin=2em]
		\item EncoderAppStatic \vspace{0.5em} \\ \vspace{0.5em}
		-c encoder\_lowdelay\_vtm.cfg \\ \vspace{0.5em}
		--InputFile=\{\textit{Input File Path}\}  \\ \vspace{0.5em}
		--InputBitDepth=8  \\ \vspace{0.5em}
		--OutputBitDepth=8  \\ \vspace{0.5em}
		--OutputBitDepthC=8  \\ \vspace{0.5em}
		--InputChromaFormat=444  \\ \vspace{0.5em}
		--FrameRate=\{\textit{Frame Rate}\}  \\ \vspace{0.5em}
		--DecodingRefreshType=2 \\ \vspace{0.5em}
		--FramesToBeEncoded=96 \\ \vspace{0.5em}
		--IntraPeriod=\{\textit{Intra Period}\}  \\ \vspace{0.5em}
		--SourceWidth=\{Width\} \\ \vspace{0.5em}
		--SourceHeight=\{Height\} \\ \vspace{0.5em}
		--QP=\{QP\} \\ \vspace{0.5em}
		--Level=6.2 \\ \vspace{0.5em}
		--BitstreamFile=\{Bitsteam File Path\} \\ \vspace{0.5em}
		--ReconFile=\{Output File Path\}
		\item DecoderAppStatic \vspace{0.5em} \\ \vspace{0.5em}
		-b \{Bitsteam File Path\} \\ \vspace{0.5em}
		-o \{Reconstruction File Path\}
	\end{itemize}

\section{Network Architechture\label{sec:net}}
Our SEVC is implemented based on DCVC-DC~\cite{dc} but focuses on exploiting additional spatial references for augmenting the contexts and latent prior.\par

\begin{figure}[tbp]
	\centering
	\includegraphics[width = 1.0 \linewidth]{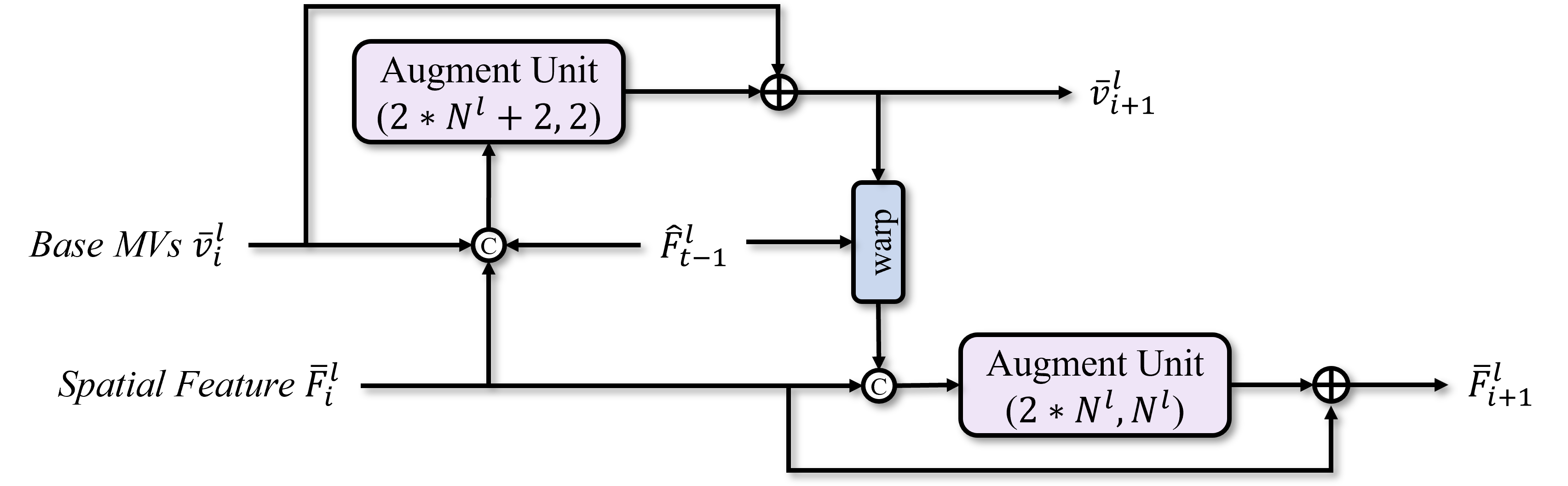}
	\caption{The network architechture of the $i$th Augment Stage.
		The numbers in an Augment Unit refer to the number of input channels and number of output channels.
		$N^l$ refers to the number of channels in the $l$th scale.}
	\label{fig:SM_1}
\end{figure}

\begin{figure}[tbp]
	\centering
	\includegraphics[width = 0.9 \linewidth]{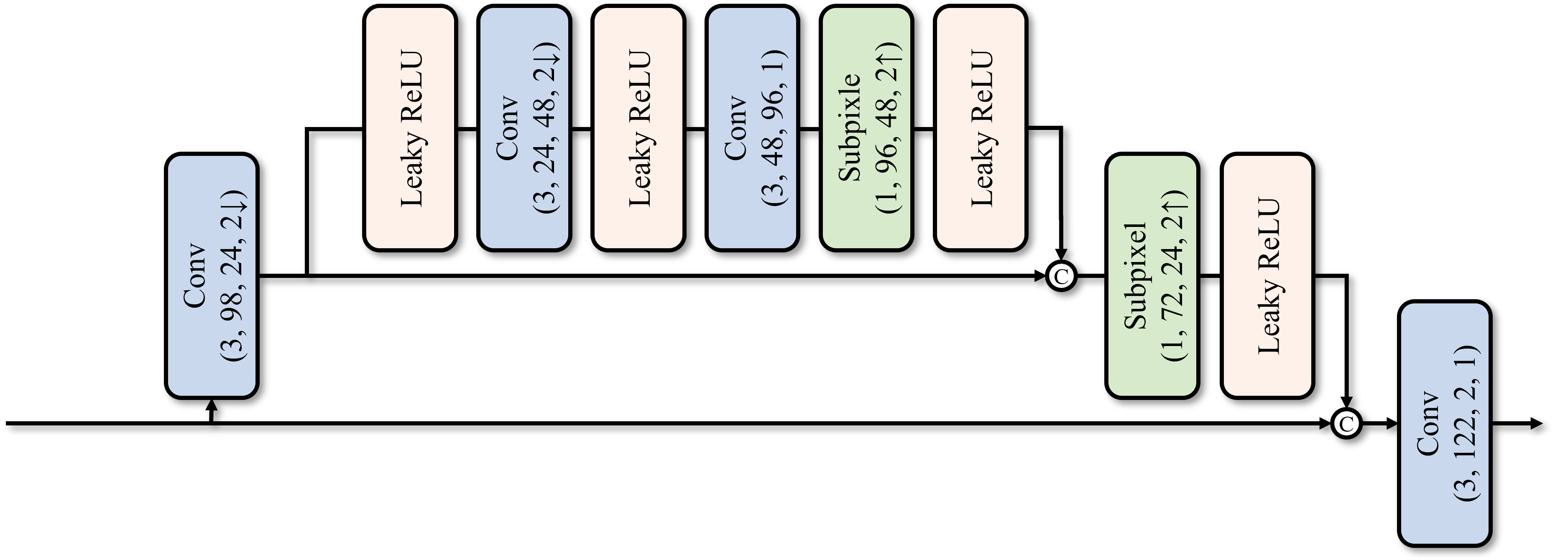}
	\caption{The network architecture of the Augment Unit.
		The numbers in a Conv block refer to the kernel size, number of input channels, number of output channels, and stride.
		This example is the Augment Unit for motion augmentation in the largest scale.}
	\label{fig:SM_2}
\end{figure}

\textbf{Motion and Feature Co-Augmentation.}
As shown in Figure~\textcolor{red}{4}, the Motion and Feature Co-Augmentation (MFCA) module
progressively improves the quality of the base MVs and the spatial feature through several Augment Stages.
Figure~\ref{fig:SM_1} shows the architecture of one Augment Stage.
It takes two steps to augment the base MVs $\bar{v}_i^l$ and spatial feature $\bar{F}_i^l$ within one Augment Stage:
Firstly, base MVs $\bar{v}_i^l$, spatial feature $\bar{F}_i^l$, and temporal feature $\hat{F}_{t-1}^l$ are fed into an Augment Unit to generate augmented base MVs $\bar{v}_{i+1}^l$.
Secondly, the augmented base MVs $\bar{v}_{i+1}^l$ leads a better alignment of $\hat{F}_{t-1}^l$ and the aligned $\hat{F}_{t-1}^l$ are fed into another Augment Unit with spatial feature $\bar{F}_i^l$
to generate augmented spatial feature $\bar{F}_{i+1}^l$.
Figure~\ref{fig:SM_2} shows the architecture of one Augment Unit.
This example is the Augment Unit for motion augmentation in the largest scale, where $N^l = 48$.
Two convolution layers with a stride equal to 2 are used to reduce the resolution and two subpixel layers~\cite{shi2016real} are used to upsample the residual back to the original resolution.\par

\textbf{Spatial-Guided Latent Prior Augmentation}
The proposed latent prior $\bar{y}_t$ is generated by adding the residual queried from multiple temporal latent representations $\hat{y}_{t-1}, \hat{y}_{t-2}, \hat{y}_{t-3}$ to the upsampled spatial latent $\hat{y}_t^b$.
To implement this, two subpixel layers are first used to upsample the spatial latent $\hat{y}_t^b$.
The upsampled $\hat{y}_t^b$ and temporal latent representations $\hat{y}_{t-1}, \hat{y}_{t-2}, \hat{y}_{t-3}$ are concatenated and fed into several Residual Swin Transformer Blocks (RSTBs)~\cite{liang2021swinir,shi2022rethinking}
to generate the residual.
Within each RSTB, there are several Swin Transformer Layers (STLs) that utilize 3D window partitions to capture correlation across the spatial and temporal dimensions.
We set the head number to 8, the window size to 8, and the embedding dimension to 128.
A Swin Transformer Layer with shifted window partitions is denoted as STL-SW.
\begin{figure}[tbp]
	\centering
	\includegraphics[width = 0.9 \linewidth]{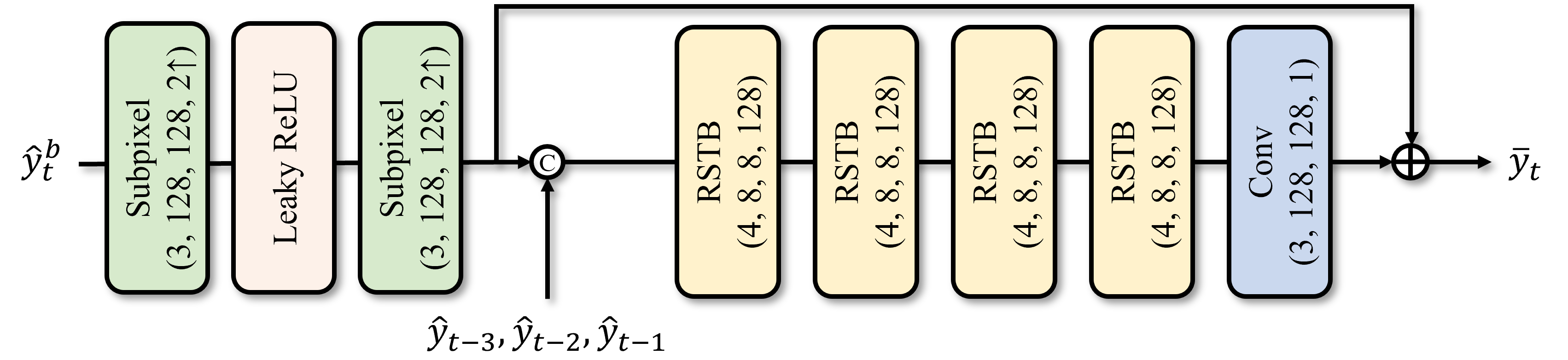}
	\caption{The network architecture of the latent prior generation.
		The numbers in a Residual Swin Transformer Block (RSTB)~\cite{liang2021swinir} refer to the depth, head number, window size, and the embedding dimension.}
	\label{fig:SM_3}
\end{figure}
\begin{figure}[tbp]
	\centering
	\includegraphics[width = 0.75 \linewidth]{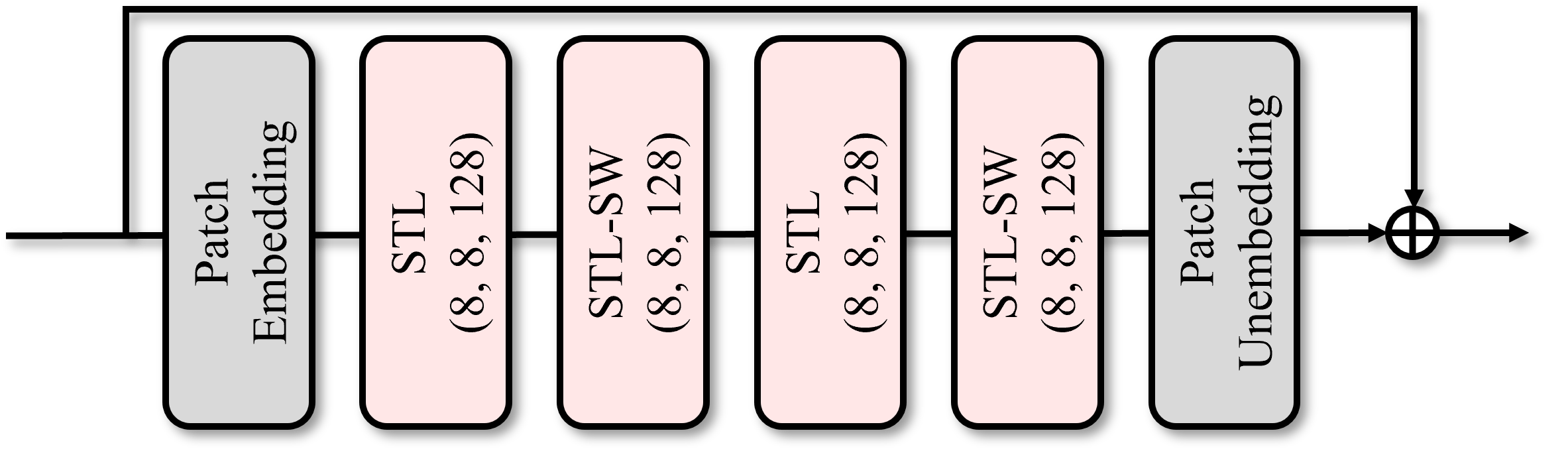}
	\caption{The network architecture of the RSTB.
		The numbers in a Swin Transformer Layer (STL) refer to the head number, window size, and the embedding dimension.
		STL-SW indicates STL with shifted window partitions.}
	\label{fig:SM_4}
\end{figure}

\begin{table}[tb]
	\centering
	\caption{Complexity Comparison.\label{tab:latent_complexity}}
	\begin{threeparttable}
		{\small
			\begin{tabular}{lcccccc}
				\toprule[1.0pt]
				$\Delta T$ & 0   & 1   & 2    & 3    & 4    \\ \midrule
				MACs       & 50G & 75G & 100G & 125G & 150G \\
				\bottomrule[1.0pt]
			\end{tabular}
			\scriptsize{
			\begin{tablenotes}
				\item[1] Tested on 1080p sequences.
			\end{tablenotes}}
		}
	\end{threeparttable}
\end{table}

In our SEVC, the Transformers are calculated on low-resolution latent representations, which will not bring too much computation cost.
Table~\ref{tab:latent_complexity} gives the complexity comparison when different numbers of temporal latent representations are used for augmentation.
It can be observed that introducing more temporal latent representations only results in a linear complexity increase.

\begin{figure*}[tb]
	\centering
	\includegraphics[width = 1.0\linewidth]{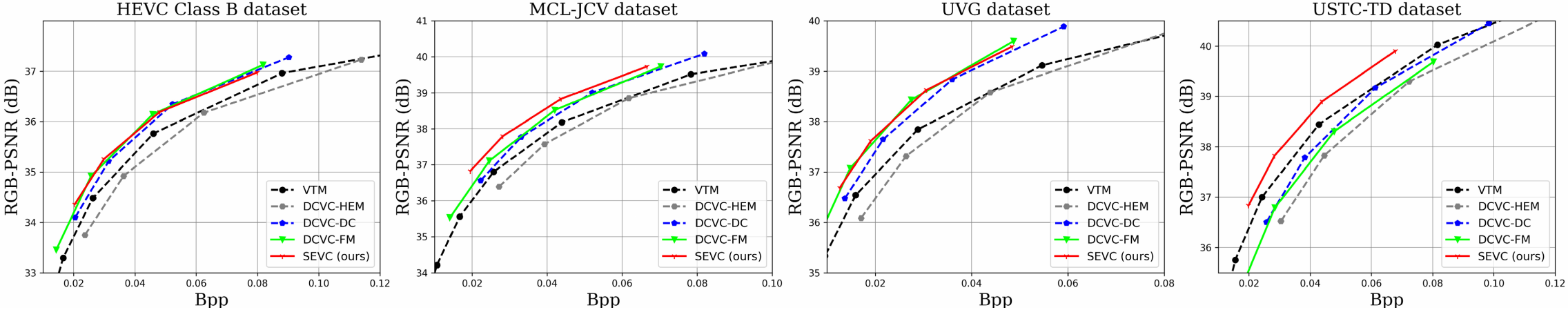}
	\caption{Rate and distortion curves on four 1080p datasets. The Intra Period is --1 with 96 frames.
	}
	\label{fig.709_rd}
\end{figure*}

\begin{table*}[tb]
	\centering
	\caption{BD-Rate (\%) comparison for RGB PSNR with BT.709. The Intra Period is --1 with 96 frames. The anchor is VTM-13.2 LDB.\label{tab:709}}
	{\small
		\begin{tabular}{lcccccccc}
			\toprule[1.0pt]
			                                   & HEVC~B      & MCL-JCV     & UVG         & USTC-TD     & Average     \\ \midrule
			\small DCVC-HEM~\cite{hem}         & 13.4        & 11.4        & 12.5        & 27.1        & 16.1        \\ \midrule
			\small DCVC-DC~\cite{dc}           & --13.4      & --13.5      & --20.6      & 11.8        & --8.9       \\ \midrule
			\small DCVC-FM~\cite{li2024neural} & \bf{--18.1} & --15.9      & \bf{--27.9} & 23.1        & --9.7       \\ \midrule
			\small SEVC (ours)                 & --16.5      & \bf{--24.5} & --27.0      & \bf{--14.5} & \bf{--20.6} \\
			\bottomrule[1.0pt]
		\end{tabular}
	}
\end{table*}

\section{Derivation of Equation~(\textcolor{red}{1})\label{sec:derivation}}
Considering downsampling the input full-resolution video $x$ to the low-resolution (LR) base video $x^b$, with their respective sources denoted as $X$ and $X^b$.
From a perspective of information theory~\cite{cover1999elements}, the mutual information between the source of $x^b$ and $x$ is
\begin{equation}
	\setlength{\abovedisplayskip}{3pt}
	\setlength{\belowdisplayskip}{3pt}
	I(X^b; X) = H(X^b) - H(X^b | X),
\end{equation}
where $X^b$ and $X$ denote the source of the base video and original video.
Given that $x^b$ is fully derived from $x$ through a fixed downsampling algorithm,
The conditional probability $p(x^b | x)$ is constant to 1.
Thereby, the conditional entropy
\begin{equation}
	\setlength{\abovedisplayskip}{3pt}
	\setlength{\belowdisplayskip}{3pt}
	H(X^b | X) = - \sum\limits_{x} p(x) \sum\limits_{x^b}p(x^b | x)\log{p(x^b|x)}
	\label{smeq:0}
\end{equation}
is constant to 0.
Therefore, we can follow
\begin{equation}
	\setlength{\abovedisplayskip}{3pt}
	\setlength{\belowdisplayskip}{3pt}
	I(X^b; X) = H(X^b).
	\label{smeq:1}
\end{equation}
Furthermore, the mutual information can be expressed equivalently as
\begin{equation}
	\setlength{\abovedisplayskip}{3pt}
	\setlength{\belowdisplayskip}{3pt}
	I(X^b; X) = I(X; X^b) = H(X) - H(X | X^b).
	\label{smeq:2}
\end{equation}
Equation (\ref{smeq:1}) and (\ref{smeq:2}) follow directly from there with
\begin{equation}
	\setlength{\abovedisplayskip}{3pt}
	\setlength{\belowdisplayskip}{3pt}
	H(X) = H(X^b) + H(X | X^b).
	\label{smeq:3}
\end{equation}
Equation~(\ref{smeq:3}) is not proposed by us, but is a conclusion well known in scalable coding and a goal that everyone wants to approach.
However, it is non-trivial to verify it for complex signals such as videos. Nevertheless, the superior performance of our SEVC makes further explorations for this conclusion.

\section{Additional Results\label{sec:add_result}}
\subsection{Results on RGB PSNR with BT.709}
When testing RGB videos, we use FFmpeg to convert YUV420 videos to RGB videos, where BT.601 is employed to implement the conversion.
However, BT.709 is used in~\cite{dc,li2024neural} for a higher compression ratio under a similar visual quality.
Thus we provide additional results with BT.709 on four 1080p datasets.
We focus on high-resolution videos because a 4x downsampling process is conducted in our spatially embedded codec,
and the spatial references with too small resolution are meaningless.\par

Figure~\ref{fig.709_rd} and Table~\ref{tab:709} give the RD curves and BD-Rate comparisons for four 1080p datasets with BT.709.
Compared to PSNR with BT.601 shown in Figure~\textcolor{red}{10}, although PSNR with BT.709 is significantly higher than PSNR with BT.601 under the same bpp,
the relative bitrate savings compared to VTM are similar.
When using BT.601 and compared to VTM, DCVC-DC achieves an average bitrate saving of 8.3\%, DCVC-FM achieves 6.2\%, and SEVC achieves 23\%. 
When using BT.709 and compared to VTM, DCVC-DC averages 8.9\% savings, DCVC-FM averages 9.7\%, and SEVC averages 20.6\%.
The slight performance degradation is attributed to the fact that the selected train set is not constructed by BT.709 conversion.

\subsection{Visualization of MVs and contexts}
In order to intuitively demonstrate the improvement in MV quality brought by our progressively augmentation,
We compare the MVs of DCVC-DC and ours using the pseudo ground truth generated by RAFT~\cite{teed2020raft}.
\begin{figure}[tbp]
	\centering
	\includegraphics[width = 1.0\linewidth]{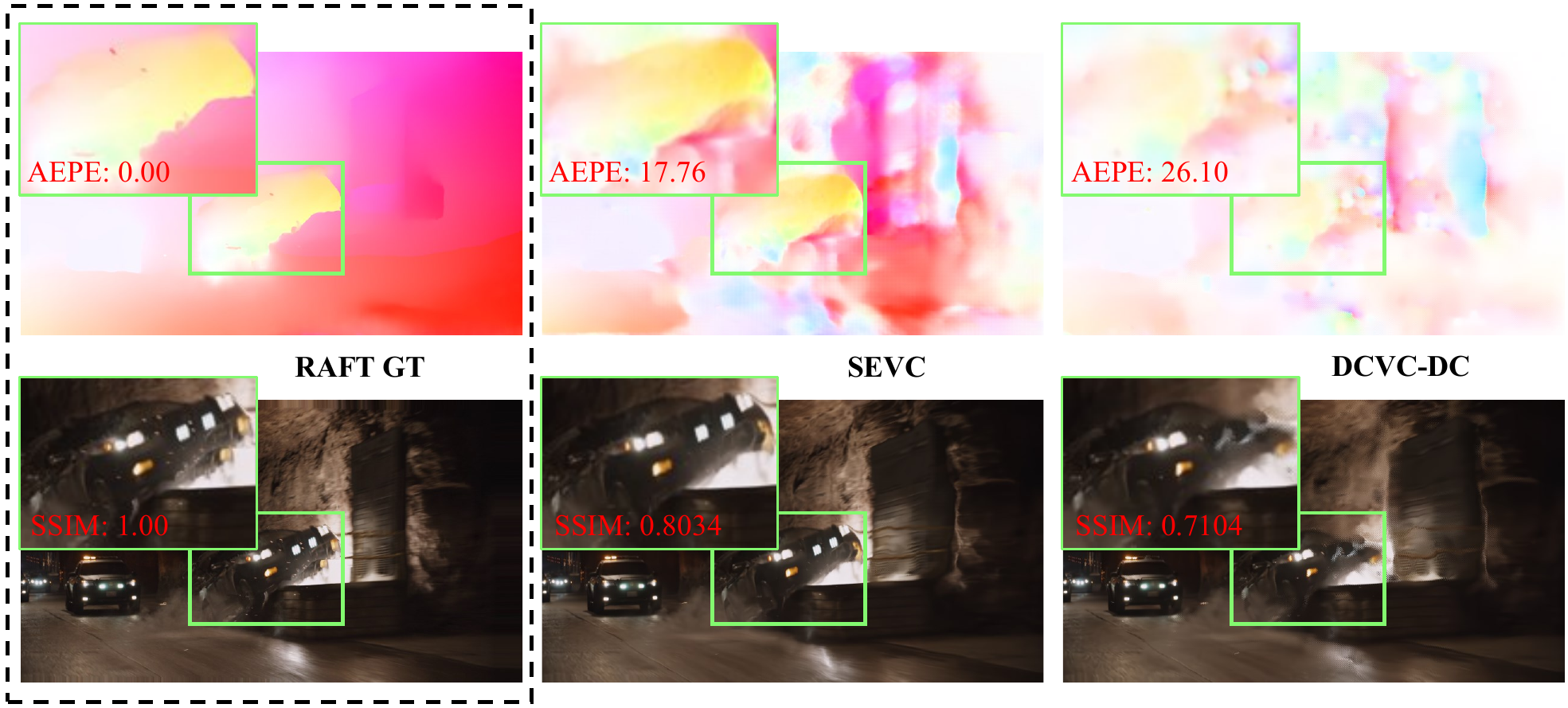}
	\caption{Comparison of reconstructed MVs and warp prediction
	of DCVC-DC and ours. Fewer AEPE scores indicate higher quality MVs and higher SSIM scores demonstrate better alignment.}
	\label{fig.mv_comparsion}
\end{figure}
Average Endpoint Error (AEPE) is used to evaluate MVs quality and Structural Similarity (SSIM) 
is used to measure warp quality of the MVs. As shown in Figure~\ref{fig.mv_comparsion}, 
both in subjective perception and objective metrics, our MV is better than that of DCVC-DC in
large motion areas whose MVs are greater than 15 pixels.

As shown in Table~\textcolor{red}{3}, our SEVC performs much better than DCVC-DC in sequences with large motions and significant emerging objects.
There are two main reasons for this:
On the one hand, the base MVs progressively augmented by our proposed MFCA module have a higher quality than the reconstructed MVs in DCVC-DC, 
which improves the utilization of temporal references.
On the other hand, the augmented spatial feature can provide an additional description for regions with emerging objects that are not well described by temporal references.\par

\begin{figure*}[tb]
	\centering
	\includegraphics[width = 1.0 \linewidth]{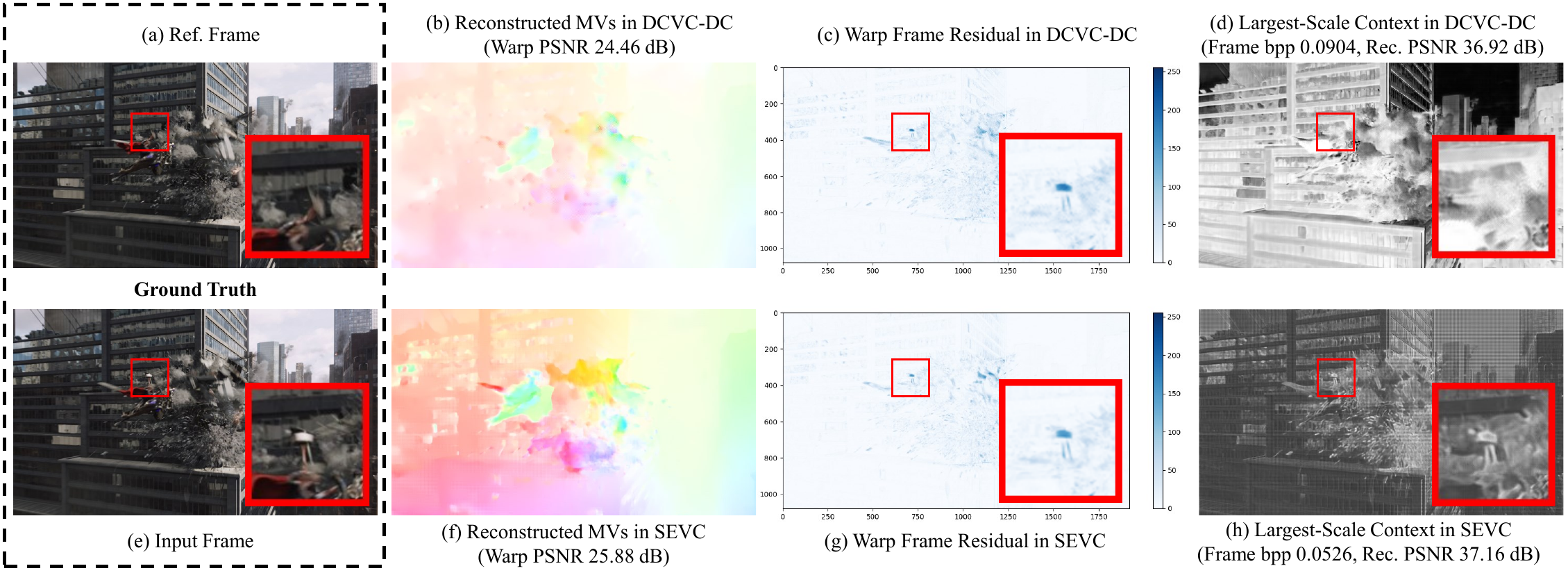}
	\vspace{-1.5em}
	\caption{
		Visualization of the MVs and contexts in DCVC-DC and our SEVC. This example is from \textit{videoSRC22\_1920x1080\_24} video of MCL-JCV~\cite{wang2016mcl}.
	}
	\vspace{-1.0em}
	\label{fig:vis_1}
\end{figure*}

\begin{figure*}[tb]
	\centering
	\includegraphics[width = 1.0 \linewidth]{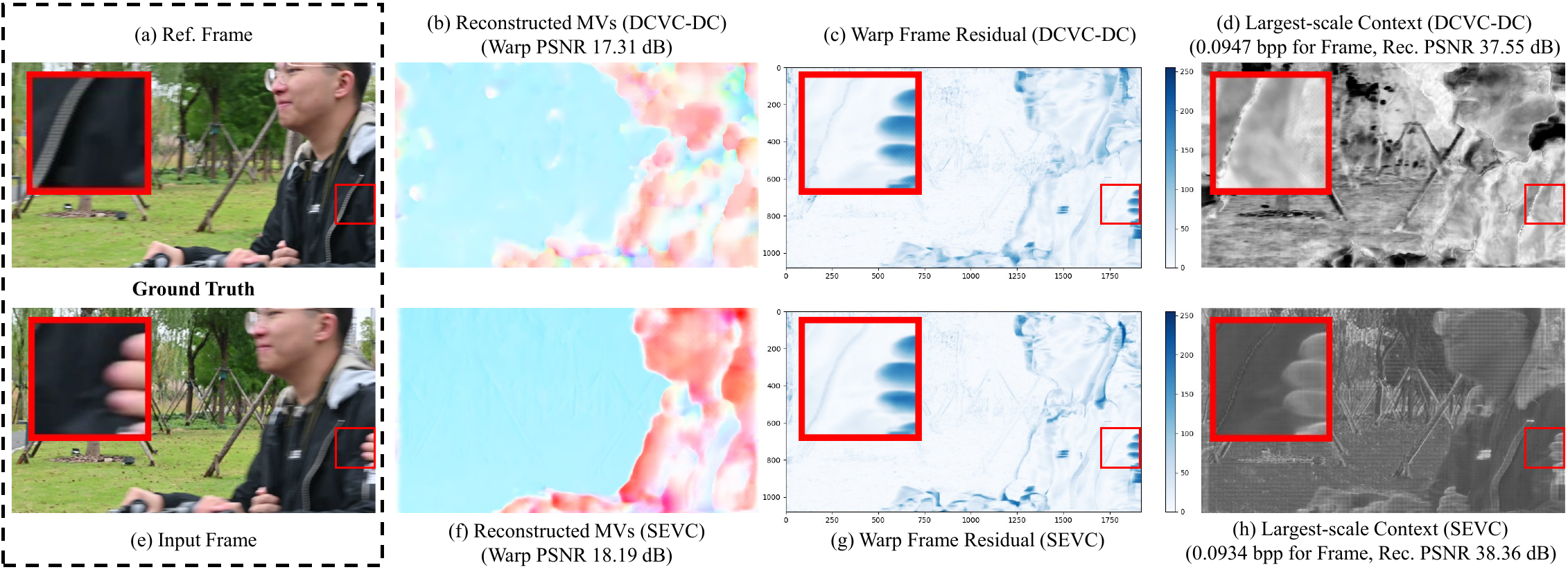}
	\vspace{-1.5em}
	\caption{
		Visualization of the MVs and contexts in DCVC-DC and our SEVC. This example is from \textit{USTC\_BycycleDriving} video of USTC-TD~\cite{li2024ustc}.
	}
	\vspace{-1.0em}
	\label{fig:vis_2}
\end{figure*}

\begin{figure*}[tb]
	\centering
	\includegraphics[width = 1.0 \linewidth]{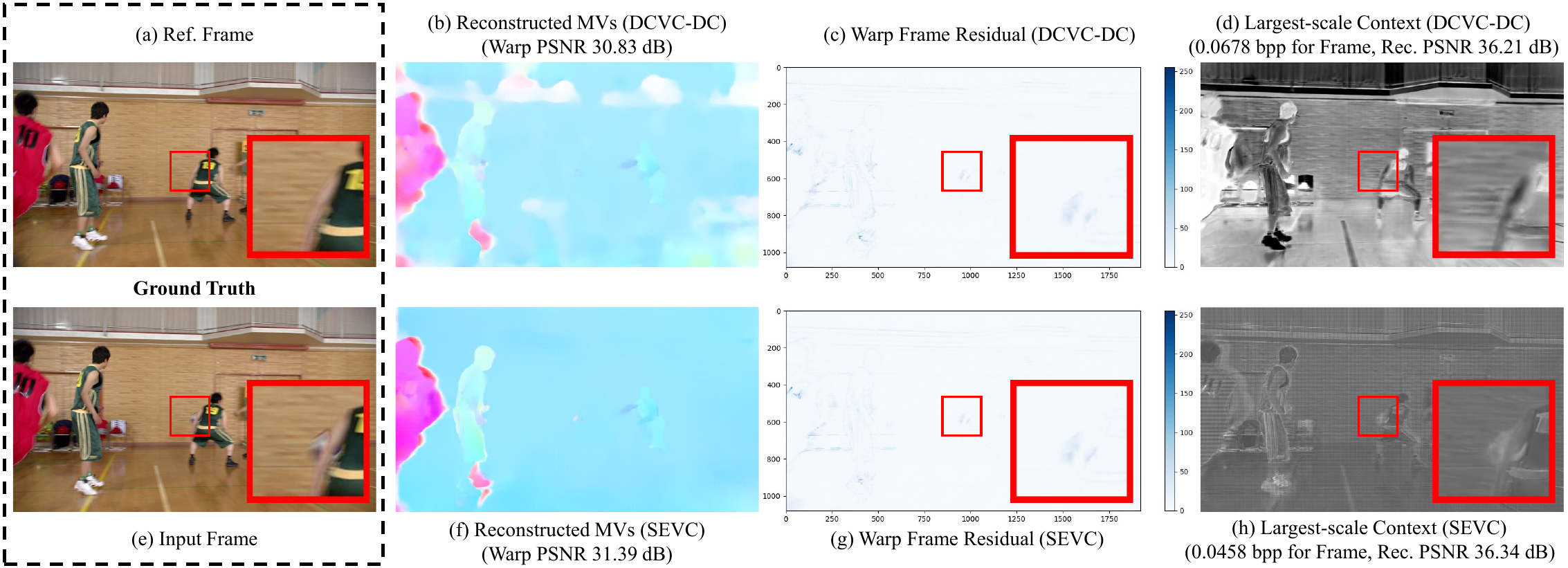}
	\vspace{-1.5em}
	\caption{
		Visualization of the MVs and contexts in DCVC-DC and our SEVC. This example is from \textit{BasketballDrive\_1920x1080\_50} video of HEVC~B.
	}
	\vspace{-1.0em}
	\label{fig:vis_3}
\end{figure*}

As shown in Figure~\ref{fig:vis_1}, Figure~\ref{fig:vis_2}, and Figure~\ref{fig:vis_3},
the augmented MVs in our SEVC have a higher warp PSNR and a better subjective quality compared to reconstructed MVs in DCVC-DC.
However, the residuals are still large in regions where new objects appear (marked in red boxes),
indicating that the temporal references are not rich enough to describe the emerging objects.
Therefore, temporal contexts in DCVC-DC fail to predict the emerging objects well.
By contrast, our SEVC utilizes an additional spatial feature, and the augmented spatial feature complements those regions.
It can be observed that in the hybrid spatial-temporal contexts, those emerging objects are well described, thus providing a better prediction.


\newpage

\end{document}